\def\etal{et al.}
\def\ie{i.e.}
\def\eg{e.g.}

\documentclass[10pt,journal,compsoc]{IEEEtran}
\usepackage{amsmath}
\usepackage{amssymb}
\usepackage{booktabs}
\usepackage{xfrac}
\usepackage{color}
\usepackage{array}
\usepackage{multirow}

\ifCLASSOPTIONcompsoc
  \usepackage[nocompress]{cite}
\else
  \usepackage{cite}
\fi
\ifCLASSINFOpdf
\else
\fi

\ifCLASSOPTIONcompsoc
  \usepackage[caption=false,font=footnotesize,labelfont=sf,textfont=sf]{subfig}
\else
  \usepackage[caption=false,font=footnotesize]{subfig}
\fi

\begin{document}
\title{Query Adaptive Late Fusion for Image Retrieval}

\author{Zhongdao Wang, 
        Liang Zheng, 
        Shengjin Wang
\IEEEcompsocitemizethanks{\IEEEcompsocthanksitem Z. Wang and S. Wang are with the 
Department of Electronic Engineering, Tsinghua University, 
Rohm Building 6-202, Beijing 100084, P.R. China.
\protect\\
E-mail: wcd17@mails.tsinghua.edu.cn, wgsgj@tsinghua.edu.cn.
\IEEEcompsocthanksitem L. Zheng is with the Research School of Engineering, Australian National University.
\protect\\
E-mail: liangzheng06@gmail.com}
}

%
%

\markboth{Manuscript}%
{Wang \MakeLowercase{\textit{et al.}}: Query Adaptive Fusion for Visual Retrieval}
%



\IEEEtitleabstractindextext{%
\begin{abstract}
Feature fusion is a commonly used strategy in image retrieval tasks, which aggregates the matching responses of multiple visual features. Feasible sets of features can be either descriptors (SIFT, HSV) for an entire image or the same descriptor for different local parts (face, body). 
Ideally, the to-be-fused heterogeneous features are pre-assumed to be discriminative and complementary to each other. However, the effectiveness of different features varies dramatically according to different queries. That is to say, for some queries, a feature may be neither discriminative nor complementary to existing ones, while for other queries, the feature suffices. As a result, it is important to estimate the effectiveness of features in a query-adaptive manner. To this end, this article proposes a new late fusion scheme at the score level. We base our method on the observation that the sorted score curves contain patterns that describe their effectiveness. For example, an ``L''-shaped curve indicates that the feature is discriminative while a gradually descending curve suggests a bad feature. As such, this paper introduces a query-adaptive late fusion pipeline. In the hand-crafted version, it can be an unsupervised approach to tasks like particular object retrieval. In the learning version, it can also be applied to supervised tasks like person recognition and pedestrian retrieval, based on a trainable neural module. Extensive experiments are conducted on two object retrieval datasets and one person recognition dataset. We show that our method is able to highlight the good features and suppress the bad ones, is resilient to distractor features, and achieves very competitive retrieval accuracy compared with the state of the art. In an additional person re-identification dataset, the application scope and limitation of the proposed method are studied.
\end{abstract}

}

\maketitle

\IEEEdisplaynontitleabstractindextext

%
\IEEEpeerreviewmaketitle

\IEEEraisesectionheading{\section{Introduction}\label{sec:introduction}}

%
%
%
%

\IEEEPARstart{T}{his} paper considers the task of
image retrieval, a critical task in computer vision. Given a query (probe) image, we
aim at searching for all the true match images in a
database (gallery). Relevant tasks include
object retrieval that searches for the
specific objects \cite{holiday,ukbench}, face/person  
recognition \cite{lfw,pipa} and pedestrian retrieval 
 \cite{market,duke,zheng2017unlabeled} that search 
for images with the same identity. 

In these image retrieval tasks, there usually exist
multiple complementary visual cues for measuring
the similarity between images. For example, in particular object retrieval, popular choices include local descriptors
like SIFT \cite{sift}, and global descriptors like GIST \cite{gist}.  For person recognition and re-identification, part-based features are often employed. 
The fusion of multiple features has been
shown to provide multiple perspectives to
describe images, pushing the retrieval accuracy
forward. Typically, if a to-be-fused feature is
moderately discriminative and complementary to
existing features, it is expected that higher
retrieval accuracy can be achieved after fusion.

There exist two mainstreams for
multiple feature fusion: early fusion and late fusion. In
early fusion, descriptors are combined at feature 
level \cite{featfusion1} or even sensor level \cite{sensorfusion1}. Then, the fused features
are processed together through the learning 
pipeline. On the other hand, late fusion refers to 
fusion at score \cite{graphfusion} or decision levels \cite{ra}. In late fusion, 
a good trade-off can be provided between the 
information content and the ease in fusion. This
paper seeks to design a simple yet effective late 
fusion scheme which works on the score level. 

A major challenge in feature fusion is that a feature's effectiveness is highly dependent on the query. Since different features demonstrate distinct strengths in finding visually similar images, global weight assignment might not be desirable when it comes individual queries. 
 For example, in particular object
retrieval, Bag of Words (Bow) based methods encode 
local texture features such as SIFT and are effective in 
finding near duplicate objects. But these methods may fail 
when the query has a smooth surface \cite{arandjelovic2011smooth}.
On the contrary, global features such as GIST and HSV 
histogram present good capability in finding globally 
similar images. But when the query has rich and distinctive texture or has occlusions and truncation, the global features might fail. 
In another example, in the task of person recognition where one needs to find the photos containing the same person with a query,   multiple visual cues could be leveraged, such as the face and clothes.
However, due to variations in pose and camera 
viewpoint, faces are sometimes invisible, and sometimes a person may even change clothes,
leading to unreliable or even harmful face/clothes features. Given
a query, one should highlight those good features and 
suppress the bad ones, otherwise the retrieval accuracy
may get even lower after fusion. In light of the above
dilemma, it is natural to raise the question how to
make an approximate estimation for the discriminative capability
of features for a specific query. This problem is not
trivial: some state-of-the-art fusion methods, as will be
shown, suffer from the inclusion of black sheep features.

Another issue that should be paid attention to includes
the amenability of the fusion method to database updating. It
requires that the fusion algorithm be independent on the 
testing database, so that its effectiveness can be preserved
in an updated database. Although offline calculations are
necessary for effective fusion, one should be aware that
an image database keeps growing, and it is desirable that
the offline steps are not dependent on it. For this issue,
some prior art requires expensive offline computations, and
the resulting systems are rigid to database change.

\begin{figure}[!t]
\centering
\includegraphics[width=\linewidth]{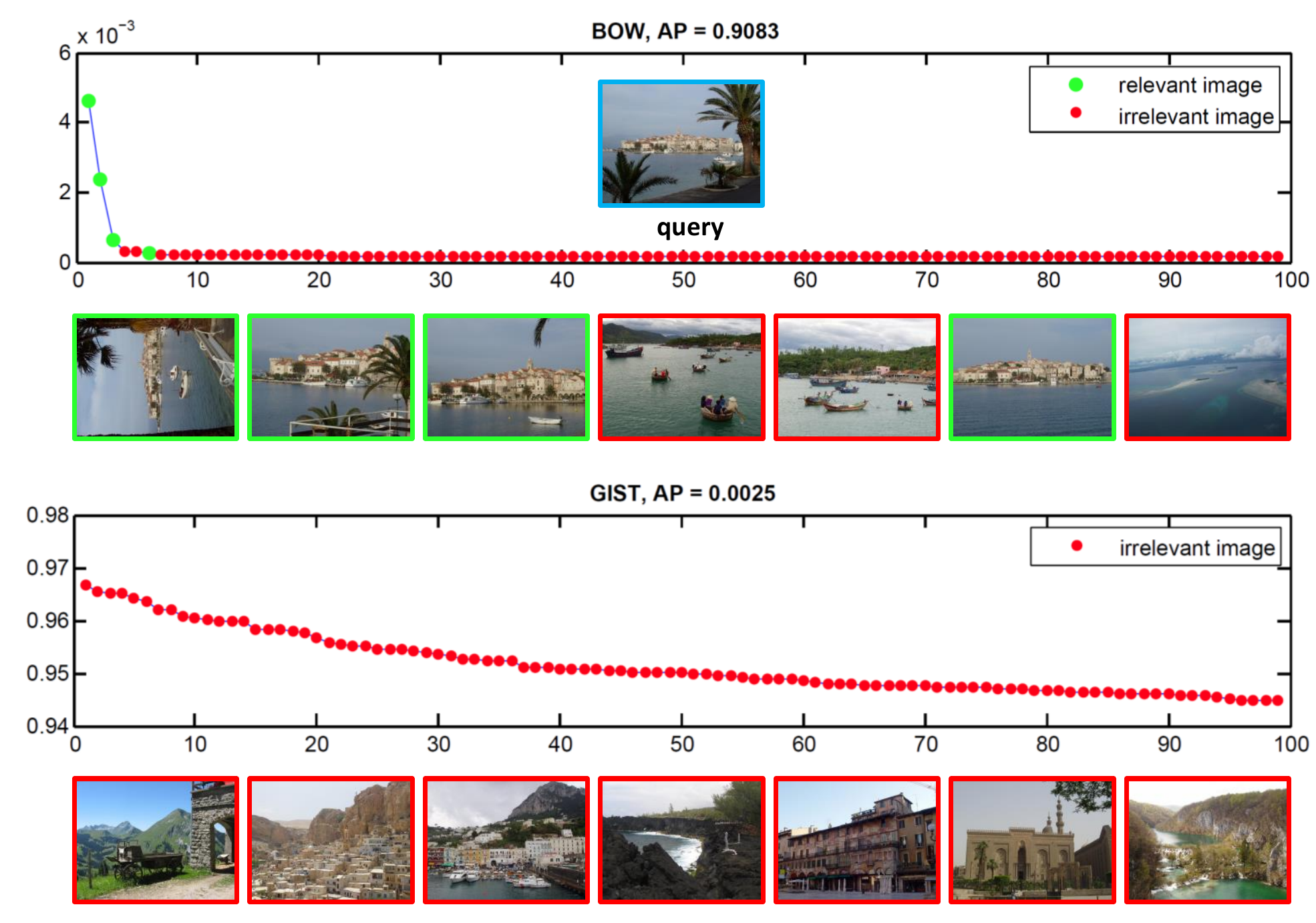}
\caption{Example of a multi-feature system. For a query 
(blue box) in the Holiday~\cite{holiday} dataset, the
SIFT (upper) and GIST (bottom) features are employed to 
obtain two score lists respectively. There are four relevant
images for this query, where SIFT produces good performance 
(AP = 90.83\%), but GIST fails (AP = 0.25\%). We plot the 
sorted scores for rank 1-99, and the corresponding 7 top-ranked 
images. True matches are marked in green, and false matches red. 
Note that the sorted score curve is L-shaped for SIFT, but descending 
smoothly and steadily for GIST.}
\label{fig:question}
\end{figure}

To address the above challenges, this paper introduces a query-adaptive late fusion method for image retrieval. The proposed method has an unsupervised (hand-crafted) version and a supervised (learning) version, both based on
a simple motivation shown in Fig. \ref{fig:question}. We observe that since good features usually generate high scores for true matches and low scores for false matches, the sorted score curve is expected 
to have an elbow point. In other words, the curve is "L" 
shaped. For bad features, the scores of true matches cannot be  distinguished from false matches, so the sorted score curve tends to be descending
smoothly and steadily.

The query-adaptive strategy is designed based on the above observation. Using this idea, we introduce two versions of query-adaptive late fusion to address different retrieval tasks, \emph{i.e.,} unsupervised and supervised. On the one hand, the unsupervised 
fusion scheme is proposed for scenarios like generic object retrieval. In this task, no prior knowledge on the topic of the query image is 
provided, since the nature of generic image search is 
unsupervised. In the proposed method, the score curves are firstly 
normalized by reference curves trained on irrelevant data, 
which are expected to approximate the tails of the initial 
score curve. Then, feature effectiveness is estimated as
inversely proportional to the area under the normalized 
score curve (see Fig \ref{fig:pipeline} for the pipeline). 
In the unsupervised fusion scheme, the offline operation is
independent on the test database, making it well suited to 
dynamic systems where the database keeps growing. More 
importantly, our method identifies ``good'' and ``bad''
features on the fly, and the results are competitive to
the state-of-the-arts on two particular object retrieval
datasets.

On the other hand, for tasks such as person recognition and pedestrian retrieval, we introduce a supervised counterpart. Under this scenario, we are provided with informative labels which are beneficial for the system. Moreover, since the image content is fixed, the variance of score curve distribution is relatively small, making it easier to fit with a parametric model. 
Experimental
results show that the query adaptive fusion module makes 
reasonable predictions of a feature's effectiveness. For instance,
in person recognition, if the query is a man
captured from the back view, our method will down-weight the
``face'' feature by assigning a low weight to it. Using the
adaptive weight for complementary features, we produce
comparable results against the state-of-the-art methods on the PIPA 
and Market-1501 datasets. 

The remainder of this paper is organized as follows.
First, we briefly review the related works in Section 
\ref{sec:relatedwork}. Then, Section \ref{sec:features} 
introduces the component features to be fused in experiment. 
We describe the query-adaptive fusion method in Section \ref{sec:method}. 
The experimental results are presented
in Section \ref{sec:exp} and conclusions are given in Section
\ref{sec:conclusion}.

\section{Related Work}\label{sec:relatedwork}

Image retrieval is a fundamental problem in computer vision. 
Among all the sub-tasks, particular object retrieval is a 
widely studied problem, aiming at searching in a database
(gallery) for all the relevant targets to a specific
query (probe) object. The target objects usually vary from
common objects \cite{ukbench} to buildings \cite{holiday, oxford, paris}.
Several descriptors present heterogeneous descriptive capability
for images. Color histogram and GIST\cite{gist} encode the holistic
images into a global feature so that images which have similar
features look alike at a glance, but the content in the images 
may be irrelevant due to the neglect of local structures.
The SIFT\cite{sift} descriptor delineates local structure
in an image and shows robustness against variations of rotation, 
brightness, and scales, but tends to be helpless when imposters
are a smooth surface \cite{arandjelovic2011smooth}. 
In order to take full advantage of the strength of each feature,
prior arts\cite{boc,cmi,coindex,graphfusion} propose various 
fusion methods. Based on the BoW structure, local features such
as color, can be combined with texture\cite{texture} or SIFT 
either by a Bag of Colors (BoC)\cite{boc} or the coupled 
Multi-index (c-MI)\cite{cmi}. Both methods work on the indexing level,
using complementary cues to filter out false positive SIFT matches.
Zhang \etal \cite{coindex} propose a co-indexing approach to 
expand the inverted index towards semantic consistency among 
indexed images. Furthermore, Rank Aggregation\cite{ra} fuses 
rank lists generated by different features by voting, which is
straightforward but easily fails when rank lists have no overlaps.
Another good practice consists in propagating the rank list 
along nodes in a graph
\cite{conf/cvpr/LiuJLC11,journals/tip/WangLTLW12,graphfusion}. 
In \cite{graphfusion}. Through link analysis on a fused graph, 
local and global rank lists are merged with equal weight. In 
\cite{journals/tip/WangLTLW12}, a graph-based learning method 
is proposed to integrate multiple modalities for visual 
re-ranking. These methods, as mentioned, are flawed in either 
of the following two aspects. First, complementary features are assumably 
employed, so there is no fall back if an ineffective feature is
integrated. Second, reranking methods such as 
\cite{graphfusion, coindex} heavily rely on the offline steps:
all images in the database should be queried and the ranking 
results are saved. This is potentially problematic if new images
are constantly added to the database, and the offline works 
should be performed all over again.

In biometric systems, multimodal features provide heterogeneous 
information which can be fused, but the properties and calculation algorithms of such features are dramatically different, giving rise
to the difficulties in feature level early fusion. As a consequence,
to compensate for the limitations in the performance of individual 
features, late fusion receives extensive attention. Nandakumar \etal
~\cite{biometric1} model the distributions of genuine and impostor 
match scores as the finite Gaussian mixture model. Jain \etal 
~\cite{biometric2} propose to transform the match scores to a 
common domain and the normalization schemes are data-dependent.
The classifier output can also be combined using a supervised 
non-Bayesian method \cite{biometric3} which minimizes classification 
error under $l_1$  constraints. For each sample, these methods 
determine a fixed weight for a specific classifier and does not 
adapt to sample variations. In~\cite{biometric4}, user-specific 
weights are used, but it requires laborious collection of training 
samples over months.

Similar to biometric recognition / retrieval, person recognition
\cite{pipa} is also a multi-feature system where different 
visual cues are expected to be fused to draw on each other's 
strength. Person recognition can be connected to face recognition \cite{lfw,megaface} and person re-identification (retrieval) \cite{lfw,megaface} \cite{market,duke}. Face recognition
typically retrieves a target person with a near-frontal face image.
Pedestrian retrieval seeks  
the target person with a holistic human body image in low resolution
(where detailed regions like faces are often not distinguishable). In comparison, the
person recognition task aims to recognize people in high resolution
images such as photos in social media photo album. 
Despite the high image resolution so that visual information like
faces and clothes can be exploited, these images present large variations in pose, clothing and camera viewpoint. Person recognition poses its unique challenge: 1) human
faces may be invisible; 2) the target person wears different clothes from the query. 
Therefore, the face / clothes features are not always reliable or even harmful for recognition. In this area, previous works usually adopt a global weight strategy. That is, the weight of face and clothes features does not change \emph{w.r.t} the query. For example, Oh \etal
 \cite{pr1}  fine-tune multiple AlexNets\cite{alexnet} with six visual 
 parts and directly concatenate the features. Zhang \etal \cite{pipa}
 and Kumar \etal \cite{pr2} train pose-specific models 
 and fuse the output similarity score either by directly summing or learning
 uniform weights for each model. Liu \etal \cite{cocoloss} propose to
 learn more discriminative part features via Congenerous Cosine Loss and 
 finally fuse similarity scores by weighted sum, in which the weights are 
 regressed in the validation set and uniform for a particular feature.
 None of the aforementioned work considers assigning query adaptive weights
 to heterogeneous features. In fact, as to be shown later, being able to assign weights in a query-adaptive manner is essential to 
 boost the retrieval performance and enhance the reliability of the
 systems.

\section{Heterogeneous Features for Various Tasks}\label{sec:features}
In this section, we provide a brief introduction to the to-be-fused heterogeneous features. A set of features can be
categorized into two types: 1) features extracted by different descriptors,
and 2) features extracted by different visual regions. Below, we will describe the heterogeneous features used in four important visual tasks. 

\subsection{Particular Object Retrieval}\label{sec:object_retrieval}
In particular object retrieval, features extracted by different 
descriptors are fused, namely BoW, HSV histogram, GIST, Random Projection and  
features from the convolutional neural network (CNN).

\begin{itemize}
    \item \textbf{Bag-of-Words (BoW).} We adopt the baseline in \cite{holiday},
and the implementation setup in \cite{cmi}. Hessian-Affine
detector and SIFT descriptor are coupled in feature extraction.
A 20k codebook is trained on Flickr60k dataset~\cite{holiday}. We
use 128-bit Hamming signature with the Hamming threshold and weighting parameter
set to 52 and 26, respectively. We also employ rootSIFT~\cite{rootSIFT}, average
IDF~\cite{idf}, and the burstiness weighting ~\cite{burstiness}. Standard inverted 
index is leveraged, and the scores are $l_2$-normalized.
    \item \textbf{HSV Histogram.} For each image, we compute an $l_2$-normalized, 
1,000-dim HSV histogram. The number of bins for H, S, V are 20, 10, 5, respectively.
    \item \textbf{GIST.} We calculate an $l_2$-normalized, 512-dim GIST~\cite{gist}
descriptor. The images are resized to $256\times256$. Four scales
are used, and the number of orientations for each scale is (8,
8, 8, 8).
    \item \textbf{Features.} Recent convolutional neural 
network based features demonstrate promising performance in image retrieval, 
integrating both local and global descriptive capability \cite{rmac, deepretrieval}. 
We adopt the state-of-the-art and publicly accessible model in \cite{deepretrieval} 
to extract an $l_2$-normalized, 2048-dim convolutional feature.
    \item \textbf{Random Projection.} To illustrate the robustness of our
method to “bad” features, we generate a random transform matrix 
$P \in \mathbb{R}^{d\times m}$ \cite{randomfeature}, where $d$ is the 
target feature dimension (set to 1000 in our experiment), and $m$ is the
dimension of the input image (with all pixels concatenated by columns).
Entries in $P$ are sampled independently from a zero-mean normal
distribution, and each row is $l_2$-normalized to unit length. In effect, 
the resulting $d$-dim feature vector $y$ is computed as $y = Px \in
\mathbb{R}^d$, where $x$ is the column-wise input image. 
\end{itemize}

\subsection{Person Recognition}\label{sec:person_recognition}
In person recognition, we employ the features extracted by different regions.
To be specific, four regions are used, \emph{i.e.}, face, head, upper body and holistic body. 

\begin{itemize}
    \item \textbf{Face.}  The head bounding boxes are given in PIPA~\cite{pipa} dataset. We 
    employ a face detector~\cite{mtcnn} to detect the accurate locations of faces inside the head bounding boxes.
    If not detected, which means the face is usually profile or even does not appear,
    a centric region in the head bounding box is extracted as the face. The faces are 
    then aligned with five facial landmarks. Finally the aligned faces are fed into a 
    20-layer ResNet and we take the output 512-dim embedding as face features.
    \item \textbf{Head.} 
    We cropped the heads according to the bounding boxes and then feed them into an 
    Inception-v3 network. The output features are 2048-dim.
    \item \textbf{Upper body.}  Given a head bounding box with up-left location $(x,y)$ 
    and size $(w, h)$, we estimate the upper body to be a box at $(x-0.7w,y)$ with size $(2.4w,3h)$,
    then the cropped image is fed into an Inception-v3 network and output the 
    feature of upper body.
    \item \textbf{Holistic body.} We estimate the holistic body to be a box at $(x-0.7w,y)$ with size 
    $(2.4w,5.5h)$, then the cropped image is fed into an Inception-v3 network and output the 
    feature of holistic body.
\end{itemize}


\subsection{Person Re-identification (Retrieval)}\label{sec:re-id}
In person re-identification (retrieval), we adopt features extracted from body regions as well, which is consistent with recent state of the art \cite{glad, pcb}. Body regions (parts) can be discovered either by unsupervised region proposals \cite{dpl, pcb} or from pose estimation results \cite{glad}. 
In this paper, we choose the Part-based Convolutional Baseline (PCB) proposed by Sun \etal \cite{pcb} for its state-of-the-art performance and simplicity in implementation. PCB is a single-branch network which is able to produce multiple part features. 

Using PCB \cite{pcb} as our feature extractor, we reproduce it and set the number of features to 6 as recommended. In this paper, \textbf{the heterogeneous features refer to those extracted from the uniformly partitioned pedestrian regions from the top to the bottom}. The similarity scores produced by different features are used for further fusion.

For particular object retrieval (Section \ref{sec:object_retrieval}), person recognition (Section \ref{sec:person_recognition}) and person re-identification (Section \ref{sec:re-id}), the experimental datasets and the performance of each individual feature will be described in Section \ref{sec:exp}.
\section{Our Approach}\label{sec:method}
\begin{figure*}[!t]
\centering
\includegraphics[width=\linewidth]{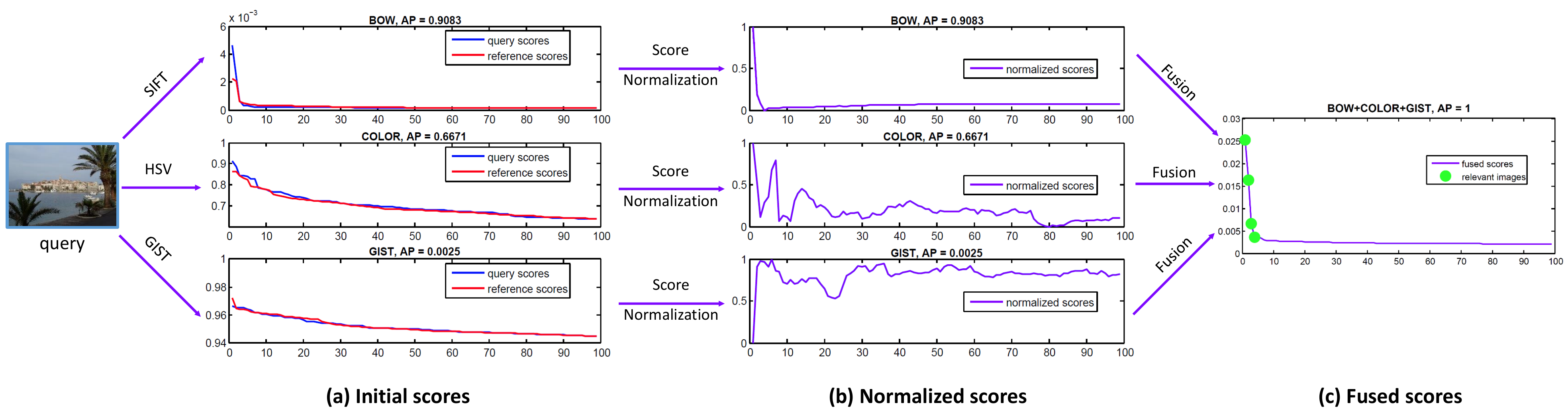}
\caption{Pipeline of the proposed 
unsupervised query adaptive fusion. Given a query image, 
three features (SIFT, HSV and GIST) are used to obtain initial rank 
lists. (a) The sorted initial scores are shown for rank 1-99 in 
blue curves, and the selected reference is depicted in 
red. (b) The tails of the score curves are eliminated by 
the reference, and the resulting scores are normalized by min-max 
normalization. (c) After calculating the feature importance through 
(b), we obtain the final score curve by Eq. \ref{eq:productrule}. 
The three features obtain APs of 0.9083, 0.6671 and 0.0025 respectively, 
and the fusion result is AP = 1.0000. The calculated query-adaptive weights are
0.69, 0.30 and 0.01 for SIFT, HSV and GIST, respectively.}
\label{fig:pipeline}
\end{figure*}

\subsection{Similarity Function}
In literature, several strategies are commonly used to combine the 
scores of multiple features in order to obtain a global confidence
measure \cite{fusionrule1,fusionrule2}, \eg, the \textit{sum, product,
maximum, minimum} rules. Among the rules, the sum rule present the 
best capability in filtering out false positives, but suffers from 
variant scales of the input scores sensitively. On the other hand, 
previous works in biometric multi-modality fusion\cite{fusionrule1, 
fusionrule2} demonstrate that the product rule has very similar, 
if not superior, performance to the sum rule. The product rule also has another
advantage: it adapts well to input data with various scales and does
not depend heavily on a proper normalization of the data. In our work,
we adopt the product rule in particular object retrieval, since
the to-be-fused features have different algorithmic procedures and thus 
the scores may vary in their scales. But for person recognition / pedestrian 
retrieval we choose to merge score lists by the sum rule because 
the to-be-fused features are generated from different human parts by 
the same algorithm, so the scales of the scores are basically the same.

Specifically, when $K$ features are fused, given query $q$ and a 
database image $d$, the similarity score of $d$ to $q$ \emph{w.r.t} feature 
$\mathcal{F}^{(i)}, i=1,...,K$ is denoted as $s^{(i)}_{d,q}$. Let 
$w^{(i)}_{q}, i=1,...,K$ encode the weight of feature $\mathcal{F}^{(i)}$ 
for query $q$, in which $\sum_{i=1}^K w^{(i)}_{q} =1$ is satisfied. Then, 
under sum and product rule, the similarity between $q$ and 
$d$ is respectively defined as,
\begin{equation}
\label{eq:sumrule}
 s_s(q,d) = \sum_{i=1}^K w^{(i)}_{q}s^{(i)}_{d,q}, 
 \text{where} \sum_{i=1}^K w^{(i)}_{q} =1,
\end{equation}
\begin{equation}
\label{eq:productrule}
 s_p(q,d) = \prod_{i=1}^K (s^{(i)}_{d,q})^{w^{(i)}_{q}}, 
 \text{where} \sum_{i=1}^K w^{(i)}_{q} =1.
\end{equation}
Note that Eq. \ref{eq:productrule} can be transformed into
a sum form by $log(\cdot)$ operator.

\subsection{Best and Worst Features}
Let us first discuss the extreme cases, \ie, the most desirable and 
the most undesirable features for a given query $q$. We assume for 
simplicity that in an image collection with $N$ images 1) there is 
only one relevant image $j^*$ to $q$ and 2) the image scores are 
normalized so that the scores range in $[0,1]$. Intuitively, the 
most desirable feature satisfies the following criteria,
\begin{equation}
\label{eq:bestcurve}
s^{(best)}_{j,q} = \left\{
\begin{array}{ll}
1,  & j=j^*\\
0,  & otherwise
\end{array} \right.
, j=1,2,...,N,
\end{equation}
where $s^{(best)}_{j,q}$ is the score of image $j$ to query $q$ \emph{w.r.t}
the best feature. In this case, only the relevant image $j^*$ receives a score of 1, and all the irrelevant images 0. To the opposite, the worst feature
for query $q$ identifies itself as assigning a score of 0 to image $j^*$ 
but 1 to the others, \ie,
\begin{equation}
\label{eq:worstcurve}
s^{(worst)}_{j,q} = \left\{
\begin{array}{ll}
0,  & j=j^*\\
1,  & otherwise
\end{array} \right.
, j=1,2,...,N.
\end{equation}
The score curves defined by Eq. \ref{eq:bestcurve} and Eq. \ref{eq:worstcurve},
once sorted, exhibit a perfect “L” and a horizontal line, respectively.
Ideally, in Eq. \ref{eq:sumrule} and Eq. \ref{eq:productrule}, weight of the 
best feature should be $w^{(best)}_{q}=1$ and that of the worst feature should be
$w^{(worst)}_{q}=0$. \textbf{We find that the weight is negatively related to the area 
under the sorted score curve.}

\subsection{Unsupervised Query Adaptive Fusion (QAF)}\label{sec:unsupervised}
We first introduce the unsupervised version of query adaptive fusion. 
In many applications such as particular object retrieval, the image content can be very diverse, making it difficult to apply large-scale 
supervised learning on each semantic categories. Also, when the features are very different in their computation algorithms, the scales of scores vary a lot across different features. Therefore, we propose to design an \textit{unsupervised} 
fusion scheme to avoid these problems. 
The proposed method is described below. 

\textbf{Reference Construction}. In our method, a critical component is the reference curve, which is used to offset the tail of a testing score curve. 
In \figurename \ref{fig:pipeline}(a), from the profiles of the three 
initial score curves, it is easy to tell that SIFT is a good feature 
for this query. But the effectiveness of HSV and GIST is not so obvious: 
both curves have a relatively ``high'' tail, and scores of the top-ranked 
images are not remarkably higher than the tail. This is expected, because 
the color and GIST are global descriptors, and there would be more images 
that share a similar global appearance with the query. In other words, 
the underlying score distribution of a feature is not considered. 

In order to alleviate the impact of ``high'' tails, this paper proposes to 
simulate the tail distribution of the initial score, by finding a reference 
score curve for each query. This reference score curve, if subtracted from 
the initial curve, would highlight the protrusion of the top-ranked scores, 
if any. In practice, we use independent datasets for reference collection. 
Specifically, for SIFT reference construction, we use the Flickr1M dataset 
released in \cite{holiday}. It contains only the SIFT descriptors, which 
is compatible with SIFT descriptors used in the test datasets. For the other 
features, we crawled 1M high-resolution images using the names of 343 countries
and regions across the world, called “Flickr343Places” dataset. Images in this 
dataset vary from scenes to objects and can be viewed as a good sampling of 
natural images.

In reference construction, we randomly select $Q$ images as queries. Then, 
particular object retrieval is performed on Flickr343Places (for HSV, CNN, 
GIST, and random features) or the Flickr1M (for SIFT) dataset. All the 
resulting image scores are sorted. Together, we collect a codebook consisting 
of $Q$ score lists for feature $\mathcal{F}^{(i)}$, denoted as 
$\mathcal{R}^{(i)}=\{r_h^{(i)}\}^Q_{h=1}$. Recall that the reference score 
lists are obtained on a dataset where all images are assumed to be irrelevant 
to each other. Therefore, the reference is able to represent the tail distribution of a score curve.

Another consideration is that the collected references should roughly be of 
the same length as the initial score curve, so that the score distribution 
would be similar. To this end, if the testing database contains $N$ images, we
should use roughly $N$ irrelevant images for reference calculation. For 
large-scale datasets where $N$ is large, both the initial score list and 
the references are down-sampled before the next step. In this manner efficiency is guaranteed.

\textbf{Query-Adaptive Feature Weighting}.
After the reference codebooks are constructed in the aforementioned offline 
method,  during the online procedure, give query $q$, the only accessible 
information we have \emph{w.r.t} feature $\mathcal{F}^{(i)}$ is the sorted score 
curve $s^{(i)}_{q}$. The profile of good feature should take on an ``L'' 
shape, while that of a bad feature a gradually descending curve (see 
\figurename \ref{fig:question} and \figurename \ref{fig:pipeline}). 

From this observation, we propose to calculate the area under the image score curve, which is taken as the indicator to feature effectiveness. As indicated above, we seek to eliminate the high tail through the subtraction by a proper reference curve. Specifically, given an initial sorted score list $s^{(i)}_{q}$ 
obtained by feature $\mathcal{F}^{(i)}$, we aim to find in $\mathcal{R}^{(i)}$ a reference which best matches the tail of $s^{(i)}_{q}$. For this strategy, the simplest method consists in finding the code in codebook $\mathcal{R}^{(i)}$ which has the smallest Euclidean distance to $s^{(i)}_{q}$, \emph{i.e.}, \begin{equation}
\label{eq:ref}
r^{(i)^*}_q = \arg \min_{r^{(i)}_h\in\mathcal{R}^{(i)}} 
\Vert s^{(i)}_{q}(u:v) - r^{(i)}_h(u:v) \Vert_2,
\end{equation}
where $h=1,2,...,Q$, and $u,v$ are parameters that restrict a curve 
segment on which the nearest neighbor is searched. Basically, it is 
required that  $v$ be relatively large to capture 
the tail distribution. Alternative to nearest neighbor search, the
tail of a score curve can also be approximated by 1) $k$-nearest 
neighbor($k$NN) search followed by an averaged sum, 2) sparse coding 
using $\mathcal{R}^{(i)}$ as the codebook. In Section 
\ref{sec:exp:imageretrieval} we will evaluate the sensitivity to parameter 
$u,v,Q$ and also the three reference search method, \emph{i.e.}, NN, $k$NN, 
and sparse coding.

\begin{figure}[!t]
\centering
\includegraphics[width=\linewidth]{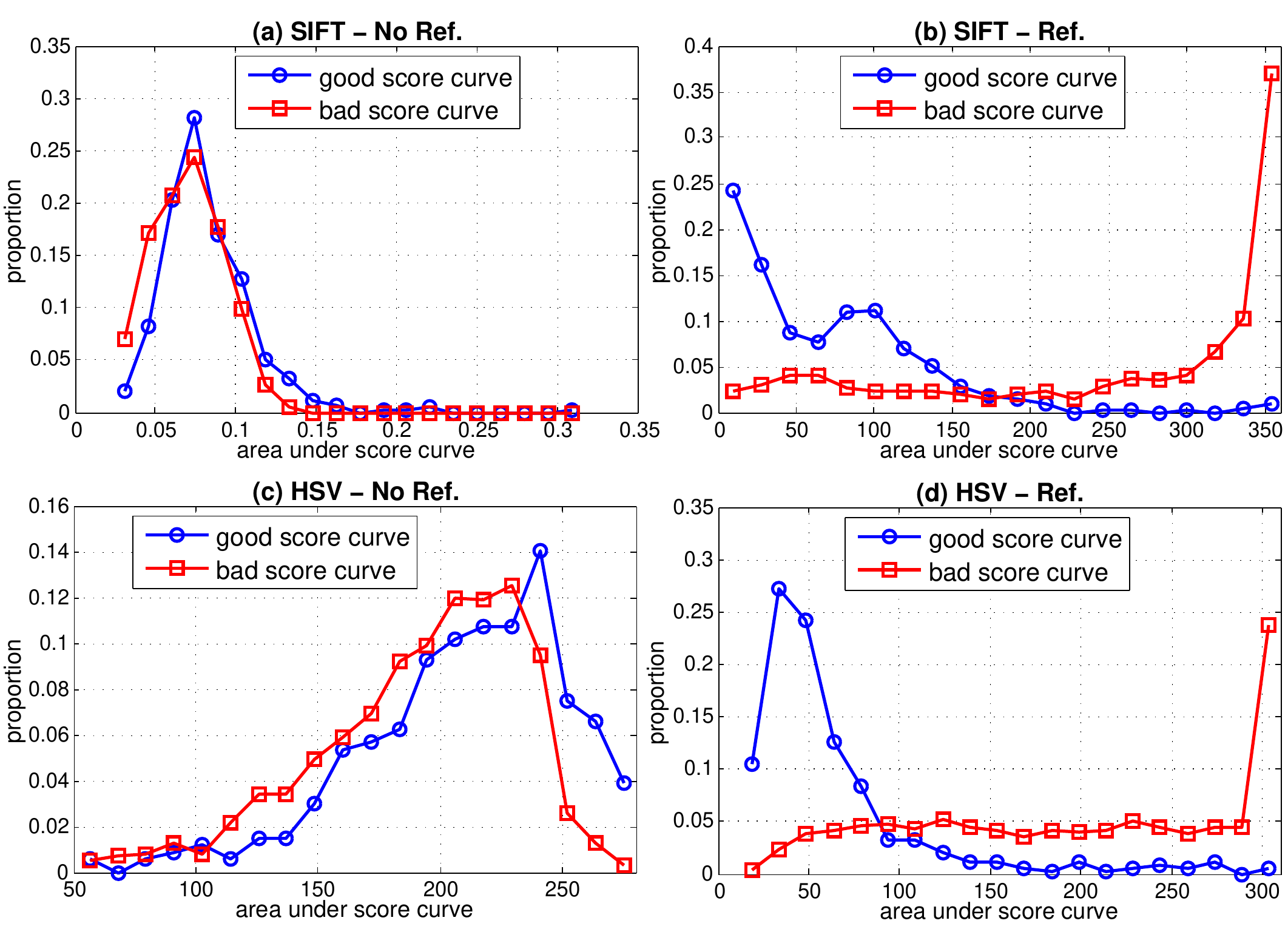}
\caption{Impact of reference subtraction. We calculate the proportion of good and bad score curves against the area under the
score curve. Without reference, for (a) SIFT and (c) HSV features,
good and bad curves cannot be distinguished. But when reference
is subtracted, for (b) SIFT and (d) HSV features, good and bad
curves are clearly separated.}
\label{fig:goodbadcurve}
\end{figure}

In the next step, the reference is substracted from the initial score curve of the query,
\begin{equation}
\hat{s}^{(i)}_{q} = s^{(i)}_{q} - r^{(i)^*}_q.  
\end{equation}
Here, as shown in \figurename \ref{fig:pipeline}(b), the reference
closely approximates the profile of the tail distribution, so that scores of the top-ranked images can be highlighted in the resulting curve $\hat{s}^{(i)}_{q}$. Subsequently, $\hat{s}^{(i)}_{q}$ undergoes min-max normalization, 
\begin{equation}
\Bar{s}^{(i)}_{q} = \frac{\hat{s}^{(i)}_{q} - \min \hat{s}^{(i)}_{q}}
{\max \hat{s}^{(i)}_{q} - \min \hat{s}^{(i)}_{q}} ,
\end{equation}
where $\Bar{s}^{(i)}_{q}$ is the normalized score curve for  
feature effectiveness estimation. To illustrate the working mechanism of reference
subtraction, for SIFT and HSV features, we have collected some good and 
bad score curves from Holidays and Flickr343Places datasets, respectively.
Good score curves are those in which rank-1 image is a true match, and bad curves are assured by the irrelevance assumption in the Flickr343Places dataset. We calculate the proportion of good and bad score curves against the area under the score curve in Fig.  \ref{fig:goodbadcurve}. 
We find that after reference  normalization, good queries tend to have a 
small area under the score curve, and vice versa. In this way, we can roughly 
tell the effectiveness of a feature after reference subtraction.

For a given query $q$ with $K$ features $\{\mathcal{F}^{(i)}\}^K_{i=1}$, 
we have $K$ score lists $\{s^{(i)}_{q}\}^K_{i=1}$. After normalization to 
$\{\Bar{s}^{(i)}_{q}\}^K_{i=1}$, the query-adaptive weight of feature 
$\mathcal{F}^{(i)}$ to $q$ is determined as,
\begin{equation}
\label{eq:weight}
w^{(i)}_q = \frac{\frac{1}{A_i}}{\sum_{k=1}^K \frac{1}{A_k}}
\end{equation}
where $A_i, i= 1,...,K$ represents the area under the $i^{th}$ feature's 
score curve. We substitute Eq. \ref{eq:weight} for Eq. \ref{eq:productrule} 
and obtain the desired query-adaptive similarity measurement.

\textbf{Discussion}.
The proposed unsupervised fusion method is featured in two aspects. First, for a given query, we estimate a feature’s effectiveness in a query-adaptive manner. While existing methods \cite{graphfusion, coindex} assign fixed weight to all 
features, our system is more robust to the impact of ineffective features. Second, by constructing a reference codebook offline, the estimation can be performed on-the-fly. Since our method does not require updating the reference codebook, and the nearest neighbor search is very fast, it can be well applied to large-scale and dynamic systems.

\subsection{Supervised Query Adaptive Fusion (S-QAF)}
In this section, we will introduce the supervised query adaptive fusion  (S-QAF), which shares the similar spirit with its unsupervised counterpart (Section \ref{sec:unsupervised}).
Unlike particular object retrieval where the content of the to-be-searched 
target is not limited, sometimes in practice, we aim to search a specific category 
of targets. Person recognition is one such task, in which the searched targets are all persons and we need to search for 
images with the same identity as the query in a database. 
The main challenges lie in the large variations in body poses, views, and consequently, 
severe deformations and even invisibility of body parts. On the one hand, in some photos, 
faces are commonly partially visible or even invisible. In such cases, the ``face'' feature 
would be unreliable, while it would be more reasonable to look at the ``body'' feature. On the other hand, 
if a person changes his / her clothes or hair-style, the ``body'' or the ``head'' features would 
be unreliable, while it would be more reasonable to look at the ``face'' feature. 
This motivates us to apply query adaptive fusion to human part 
features. In spite of the effectiveness and generalization of the aforementioned 
unsupervised query adaptive fusion (Section  \ref{sec:unsupervised}), we propose a supervised version for such 
content-specific retrieval tasks. The reason why we utilize supervised learning is two-fold. 
First, the informative labels are can be utlized, making it possible to learn to generate the optimal adaptive weights for each feature in an end-to-end manner. Second, the input content 
of the feature extractors is limited, so the distribution of the feature is limited. Consequently, 
the distribution of the similarity scores is also not as ``diverse'' as in particular object retrieval, 
making the patterns of ``good'' features and ``bad'' features embodied in the score curves more
distinct. Below, we will describe the supervised version from two aspects, \emph{i.e.}, the model and the optimization objective. 

\begin{figure}[t]
    \centering
    \includegraphics[width=\linewidth]{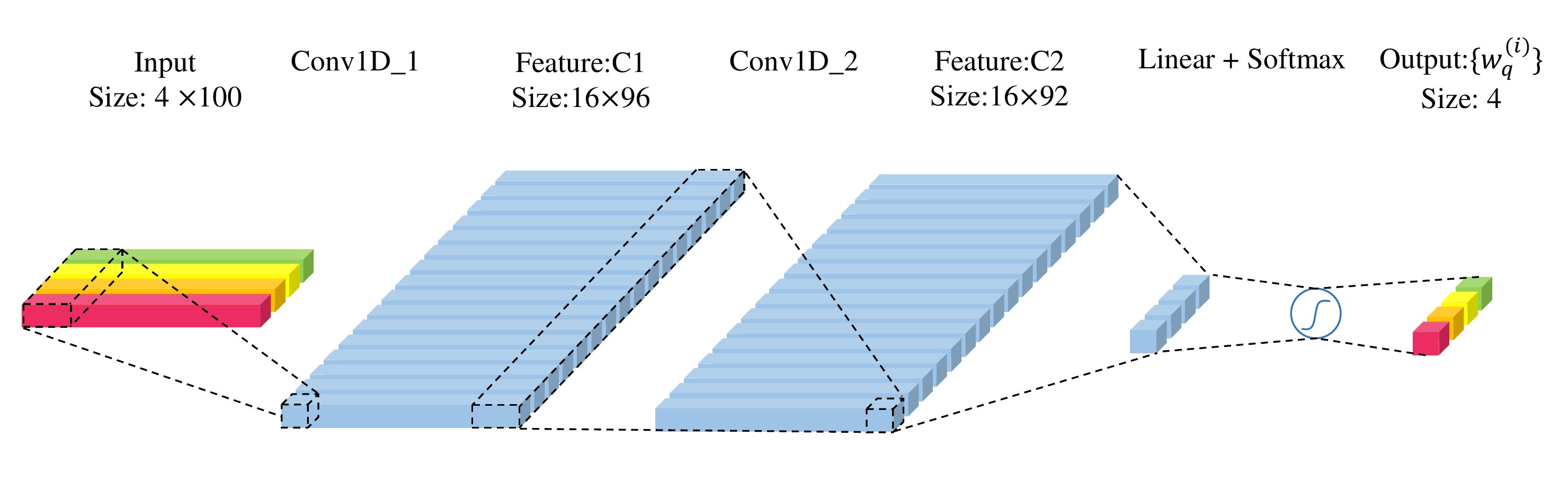}
    \caption{The network structure used in the proposed S-QAF method for person recognition (number of heterogeneous features $K=4$). For all the convolutional filters, kernel size is $5\times1$, and stride is 1.}
    \label{fig:network}
\end{figure}

\textbf{Model}. For a specific query, the query adaptive fusion module takes as input multiple 
sorted score curves and outputs the corresponding weight for each features. For
instance, in person recognition, we extract four part-based features from a person, \emph{i.e.},
face, head, upper body and holistic body. For each feature, we compute the cosine similarity 
between the query and the whole database, yielding a similarity vector $s^{(i)}_{d,q}$. 
Once sorted, the similarity vector $s^{(i)}_{d,q}$ presents a descending curve, and we 
denote the sorted score curve as $s^{(i)}_{q}$. The sorted score curves can be regarded 
as 1-D signals, from whose magnitude and differentials we can infer its effectiveness for the query. 
We observe that the top $m$ scores of $s^{(i)}_{q}$ contain 
most of the information, because the tail of the curve is close to a constant level. Therefore,
the top $m$ of the score curve, denoted as $s^{(i)}_{q}(1:m)$, is cut off as the input of 
the model. Score curves of different features are stacked along an extra dimension, similar to the
``channel'' dimension in images, and we denote the stacked curves as $S^m_{q}$, $S^m_{q}= 
[s^{(1)}_{q}(1:m), ..., s^{(K)}_{q}(1:m)]^{T}$. Given $S^m_{q}$ as input, we train 
a small neural network $f(S^m_{q};\theta)$ parameterized by $\theta$. We utilize 1-D convolutional filters to process the input to better capture of the differentials of the signal. In person recognition experiment, we set the number of features  $K$ to $4$, and set $m$ to $100$. The network structure
and input / output size is presented in \figurename \ref{fig:network}. The output of
$f(S^m_{q};\theta)$ is a vector of size $K$ which is summed up to 1.  It is expected to approximate the optimal $w^{(i)}_{q}$.

\textbf{Optimization Objective}. A brute-force solution to generating query adaptive weights is 
to manually define or search the optimal weights for features of a specific query, and use them as 
the supervision signals to train $f(S^m_{q};\theta)$. However, it must not be optimal if we manually
define the ``ground truth'' weights, since we could not quantify the importance of features accurately. 
Searching for optimal weights may be possible, but it needs a large amount of computation. Therefore, we 
propose to train $f(S^m_{q};\theta)$ in an end-to-end manner.

The basic idea is directly optimizing the gap between the average score of true matches in the fused
score curve and that of the false matches. Denoting the output of the network as $f(S^m_{q};\theta) = 
(w^{(1)}_q, ..., w^{(K)}_q)$, we use them to replace the weights in Eq. \ref{eq:sumrule} and obtain 
the unsorted fused score $s^{f} = \sum_{i=1}^K w^{(i)}_q s^{(i)}_{d,q}$. According to the labels 
of each candidate in training database and the label of the query, we categorize the scores in $s^{f}$
into a true match set $\mathcal{M^+}$ and a false match set $\mathcal{M^-}$, \emph{i.e}, for each $s^{f}_i$
in $s^{f}$, $s^{f}_i \in \mathcal{M^+}$ if $y_i = y_q$, and $s^{f}_i \in \mathcal{M^-}$ if not
$y_i = y_q$, where $y_i$ and $y_q$ are the labels of the $i$th database image and the query image,
respectively. The optimization objective can be formulated as,
\begin{equation}
\label{eq:loss}
    \mathcal{L}(s^{f}) =  \max({\frac{1}{|\mathcal{M}^-|}\!\!\sum_{s^{f}_i\in\mathcal{M^-}} \!\!(d +s^{f}_i) - \frac{1}{|\mathcal{M}^+|}\!\!\sum_{s^{f}_i\in\mathcal{M^+}} \!\!s^{f}_i},0),
\end{equation}
where $d$ is a margin term. We set $d=1$ in all our experiment. 

In practice, we also find it effective to mine \emph{hard negative matches}. We sort scores of false 
matches in $\mathcal{M^-}$ in descending order and pick $N^-_q=\alpha N^+_q$ false matches with the top
scores as hard
negative matches, where $N^+_q$ is the number of true matches for the specific query, and $\alpha$ is a
balancing ratio. Instead of calculating the average score of all false matches as in Eq. \ref{eq:loss}, we only calculate the average score of hard negative matches. Mining 
hard negative matches makes the model converge to a better point, because the ``easy'' negatives dominate 
in number and would always yield low score once averaged. We set the ratio $\alpha = 2$ in our experiment. 
\section{Experiment}\label{sec:exp}

\begin{figure}
    \centering
    \includegraphics[width=\linewidth]{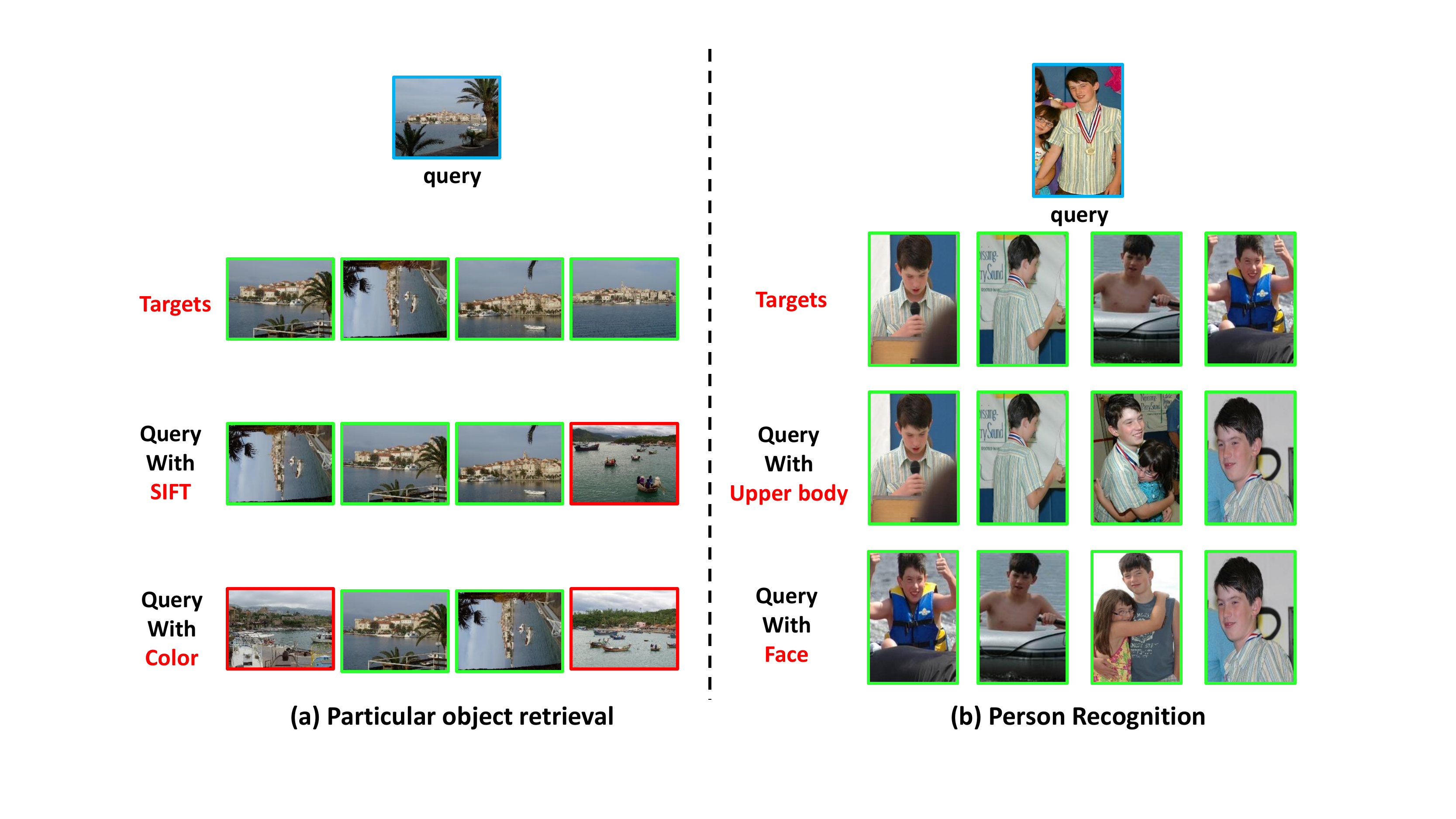}
    \caption{Illustration on (a) \textit{particular object retrieval} task and (b) \textit{person recognition} task. We can observe that different visual cues have different discriminative ability.}
    \label{fig:task}
\end{figure}
We first illustrate the tasks of particular object retrieval and person recognition in 
\figurename \ref{fig:task}. The goal of both tasks is to find the relevant targets to a
query image in a gallery set, while the specific targets of each task are the particular object and particular person respectively. As shown in \figurename \ref{fig:task}, 
different visual cues (features) have different discriminvative power, we show the merits of QAF and S-QAF by fusing these heterogeneous features.
\begin{table*}[!t]
\caption{Results on two object retrieval benchmarks using different fusion methods. We compare our method with Rank 
    Aggregation (RA)\cite{ra}, Graph Fusion (Graph)\cite{graphfusion}, and global weight tuning (Global), respectively.  
    BoW with Hamming Embedding \cite{holiday} is used. Note that Graph fusion is sensitive to parameters, to be shown in Fig. \ref{fig:graphfusion}.}
    \centering
    \begin{tabular}
       {p{1cm}<{\centering}p{1cm}<{\centering}p{1cm}<{\centering}p{1cm}<{\centering}p{1cm}<{\centering}|p{1cm}<{\centering} p{1cm}<{\centering}p{1cm}<{\centering}p{1cm}<{\centering}|p{1cm}<{\centering}p{1cm}<{\centering}p{1cm}<{\centering}p{1cm}<{\centering}}
    \toprule
    \multicolumn{5}{c|}{Features }&\multicolumn{4}{c|}{Holidays} & \multicolumn{4}{c}{Ukbench}\\ 
    \hline
    CNN& Bow& HSV& GIST& RAND& {RA}&{Graph}&{Global}&{ours}&{RA}&{Graph}&{Global}&{ours}\\
    \hline
    \hline
                  &   \checkmark&             &   \checkmark&             &
          52.00   &    76.39    &   81.54     &    80.88    &
          2.429   &    3.674    &   3.610     &    3.590                    \\
                  &   \checkmark&             &             &   \checkmark&
          39.42   &    76.57    &   81.18     &    80.91    &
          2.134   &    3.680    &   3.603     &    3.596                    \\
                  &   \checkmark&             &   \checkmark&   \checkmark&
          51.48   &    70.59    &   81.65     &    81.47&
          2.452   &    3.654    &   3.634     &    3.590                    \\
                  &   \checkmark&   \checkmark&             &             &
          83.79   &    71.98    &   84.18     &    84.47&
          3.538   &    3.808    &   3.745     &    3.755                    \\
        \checkmark&             &             &   \checkmark&             &
          52.34   &    81.61    &   93.32     &    92.96    &
          2.385   &    3.851    &   3.835     &    3.833                    \\
        \checkmark&             &             &   \checkmark&   \checkmark&
          51.35   &    81.20    &   93.33     &    93.12&
          2.429   &    3.831    &   3.835     &    3.826                    \\
        \checkmark&             &             &             &   \checkmark&
          38.43   &    80.81    &   92.25     &    93.19&
          2.114   &    3.873    &   3.835     &    3.833                    \\
        \checkmark&   \checkmark&             &             &             &
          88.12   &    91.94    &   93.66     &    93.67&
          3.788   &    3.940    &   3.910     &    3.887                    \\
        \checkmark&             &   \checkmark&             &             &
          75.03   &    86.64    &   93.70     &    93.97&
          3.754   &    3.906    &   3.912     &    3.845                    \\
        \checkmark&   \checkmark&   \checkmark&             &             &
          91.36   &    92.61    &   94.17     &    94.08&
          3.870   &    3.945    &   3.920     &    3.902                    \\
        \checkmark&   \checkmark&   \checkmark&   \checkmark&   \checkmark&
          82.33   &    86.56    &   94.20     &    94.25&
          3.626   &    3.935    &   3.920     &    3.901                    \\

    \bottomrule
    \end{tabular}
    
    \label{tab:imageretrieval}
\end{table*}

\subsection{Evaluation on Particular Object Retrieval}
\label{sec:exp:imageretrieval}
\textbf{Datasets}. We evaluate the proposed \emph{unsupervised} fusion method on two 
particular object retrieval datasets, \emph{i.e.}, Ukbench \cite{ukbench} and 
Holidays \cite{holiday}. The Ukbench dataset contains 10,200 images composed 
of 2,550 groups, each group has 4 images which are relevant to each other. 
Each image is taken as query in turn. We use N-S score as measurement, 
It counts the number of relevant images in the top-4 ranked images. 
The Holidays dataset is released with 1,491 personal holiday pictures. 
There are 500 queries in total. Mean Average Precision (mAP) is 
used as evaluation metric. It is the mean value of Average Precision 
(AP), which encodes the area under the precision-recall curve for 
each query. 

\textbf{Performance of individual features}. Table \ref{tab:individual} 
showcases the performance of individual features, \ie, SIFT encoded by BoW, 
Deep CNN, HSV, GIST, and random projections, on the two 
datasets. As we can observe, the up-to-date deep CNN based feature 
\cite{deepretrieval} presents satisfying performance on both datasets. 
Because the content of the images in Holidays are mainly scenary and landmarks, the high performance on Holidays greatly benefits from supervised training with  
a large-scale landmark dataset. On Ukbench, the fine-tuned CNN feature is reported to be less effective 
due to the inconsistency between training and testing data \cite{deepretrievaljornal}. But the CNN feature  
 still outperforms hand-crafted features. Given the individual features, we will demonstrate that 
the proposed fusion scheme is able to further boost the retrieval accuracy.

\begin{table}[!t]
\caption{Retrieval accuracy with individual features.}
    \centering
    \begin{tabular}{|l||ccccc|}
    \hline
        Datasets               &     CNN &   BoW &  HSV &   GIST &  RAND \\
    \hline
        Holidays, \textit{mAP} &  92.99  &  80.16& 61.32&  33.81 &  13.49 \\
    \hline
        Ukbench, \textit{N-S}  &  3.833  &  3.582& 3.195&  1.856 &  1.422 \\
    \hline
    \end{tabular}
    \label{tab:individual}
\end{table}

\begin{figure}[!t]
\centering
\includegraphics[width=\linewidth]{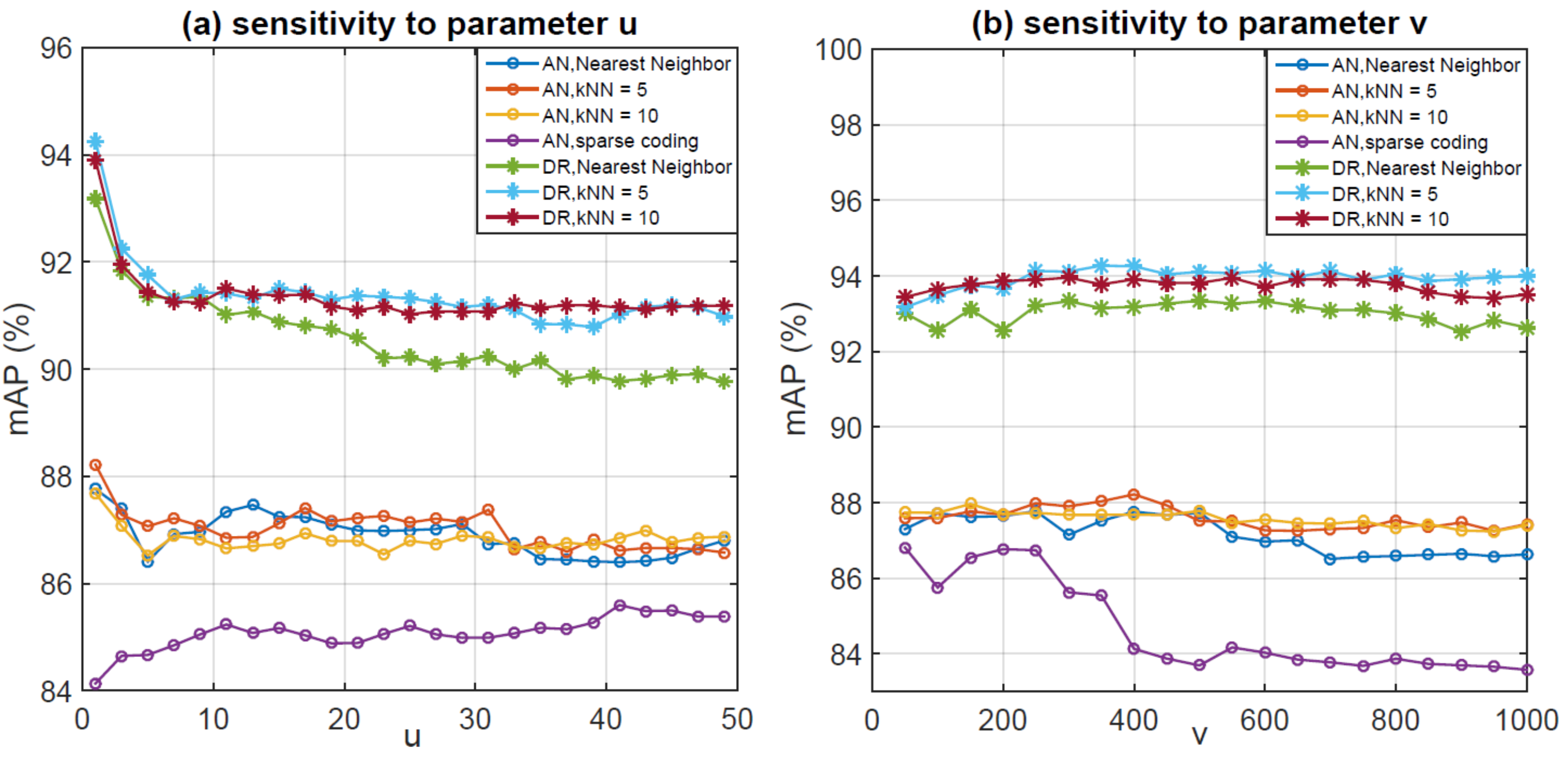}
\caption{Sensitivity to $u$ and $v$ on Holidays. Five features are fused. 
\emph{i.e.}, CNN, BoW, HSV, GIST and Random Projection. mAP is plotted against the 
two parameters in Eq. \ref{eq:ref}. Four reference selection methods are 
compared, \emph{i.e.}, nearest neighbor, $k$NN ($k = 5$ or $10$), and sparse coding. 
We also explore different CNN features, \emph{i.e.}, fine-tuned CNN feature trained with Deep 
Retrival~\cite{deepretrievaljornal} (DR) and CNN feature extracted from an 
ImageNet pre-trained AlexNet (AN).}
\label{fig:param:uv}
\end{figure}

\begin{figure}[!t]
\centering
\subfloat{\includegraphics[width=0.5\linewidth]{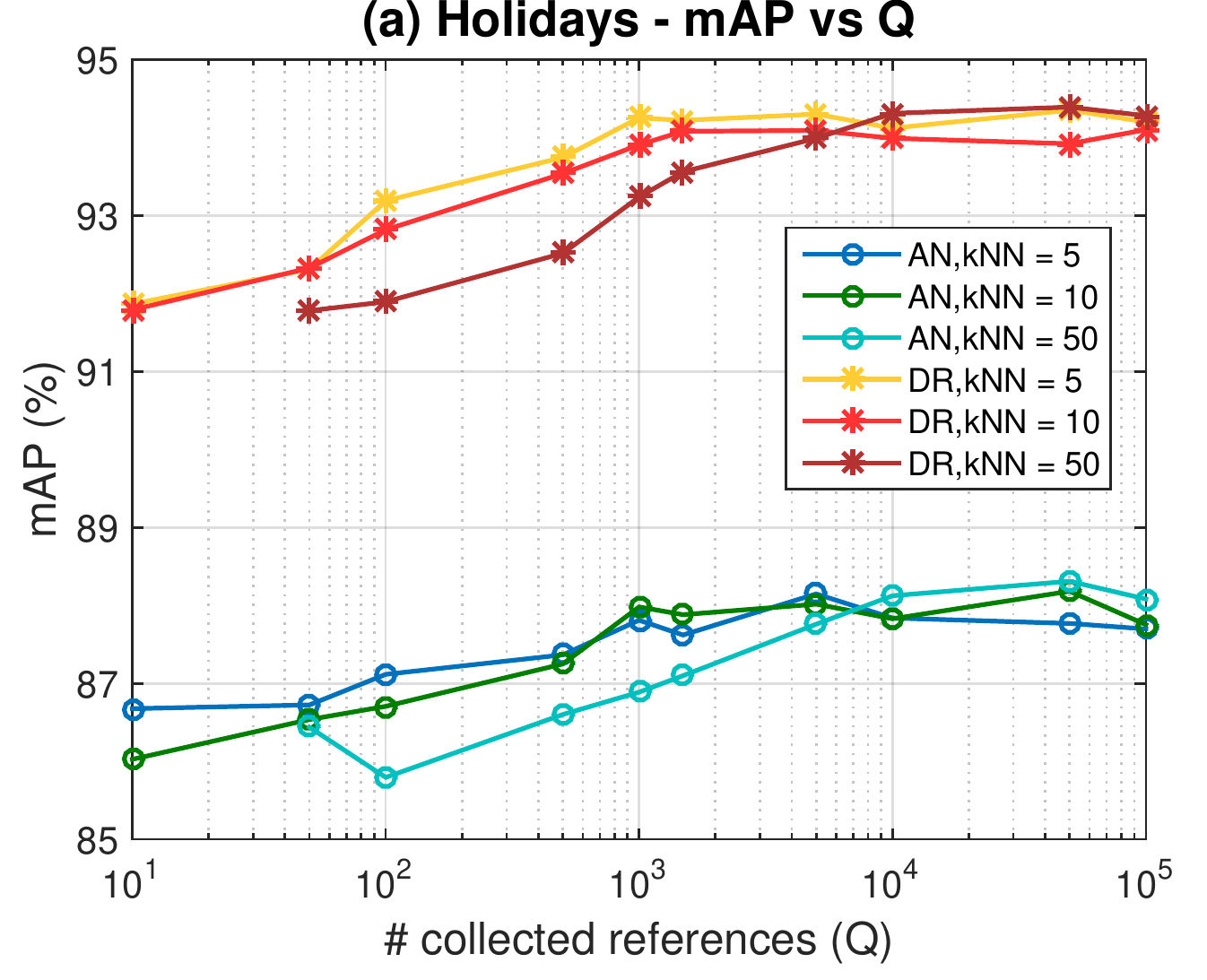}}
\subfloat{\includegraphics[width=0.5\linewidth]{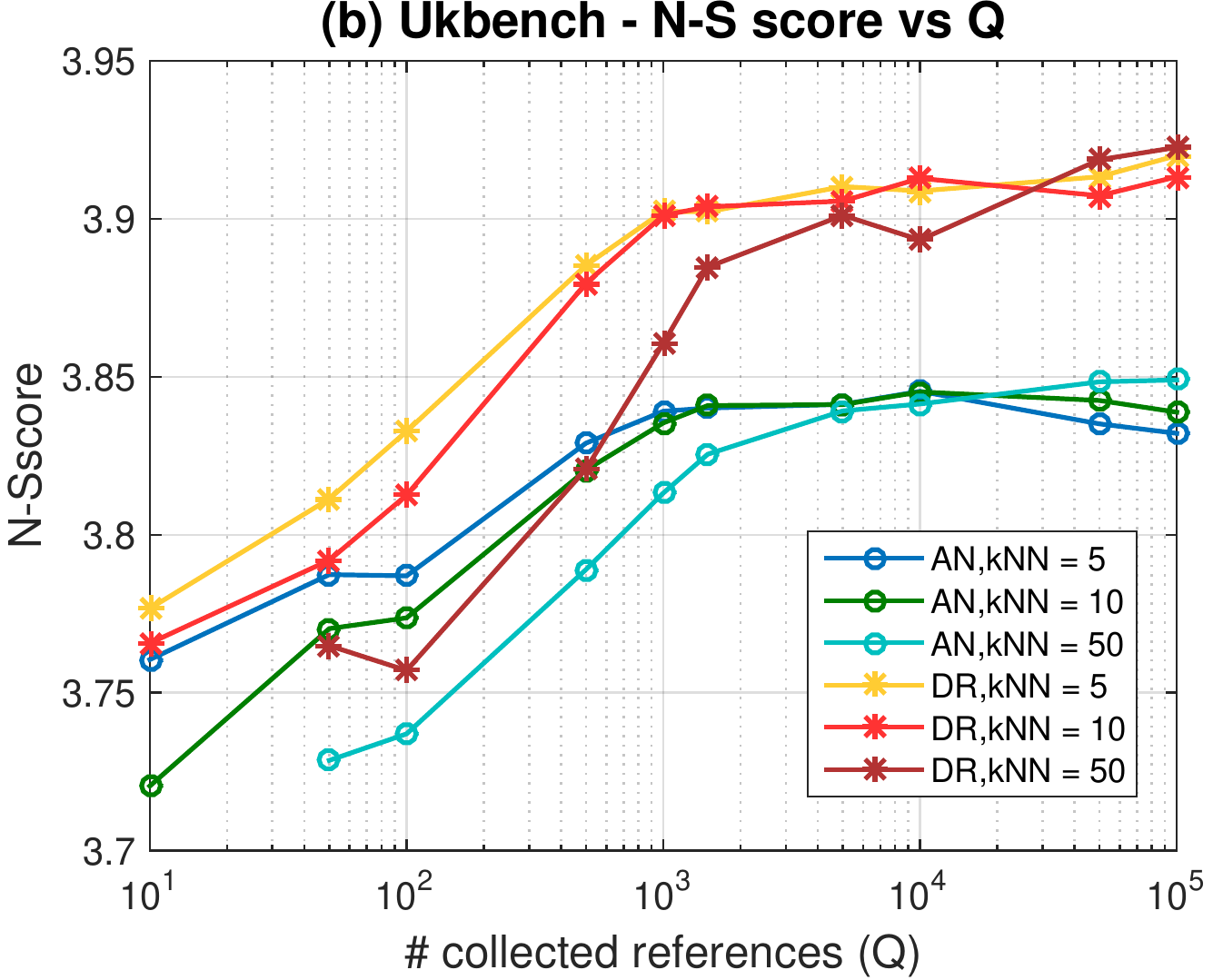}}

\caption{Sensitivity to parameter $Q$ on (a) Holidays and (b) Ukbench
datasets. We test $k$NN = 5, 10, and 50. \textit{DR} refers to Deep Retriveal 
as CNN feature extractor while \textit{AN} refers ImageNet Pre-trained AlexNet 
as CNN feature extractor.}
\label{fig:param:q}
\end{figure}

\textbf{Parameter selection}. 
Three parameters are involved in unsupervised QAF. We first evaluate the matching 
parameters $u,v$ in Eq. \ref{eq:ref}, and the results are demonstrated in 
Fig. \ref{fig:param:uv}. We employ two types of CNN features here 
for comparison, one is the discriminatively learned deep retrieval model \cite{deepretrieval} (DR), 
and the other is an ImageNet pre-trained AlexNet (AN). The rest of the to-be-fused 
features are BoW, HSV, GIST and Random Projection. We observe that,
as $u$ and $v$ vary, mAP is relatively stable for features that contains AN. For features 
that contains DR, $u=1$ is the best choice. We speculate that it is due to the peaky 
distribution of the scores generated by DR. In other words, the score is either very 
close to $0$ or very close to $1$, so a small $u$ is needed for highlighting the top scores. 
Performance of sparse coding is inferior, because the sparse control item has negative 
impact on the NN search item. Moreover, three NN-based methods perform similarly, and it 
seems that ``$k$NN = 5'' is slightly superior. When using $k$NN, the averaged reference is 
more resistant to noise, but when $k$ increases, more ``bad'' references are introduced 
especially under small $Q$. We set $u=1$ and $v=400$ in our experiment.

When evaluating parameter $Q$, \emph{i.e.}, the size of the reference codebook, we present the 
results in Fig. \ref{fig:param:q}. We find that the fusion accuracy increases 
steadily with $Q$. In fact, when $Q$ is large,
it is more likely to find among them a good approximation to the tail distribution. 
Nevertheless, computational complexity also increases with $Q$. Considering this, we 
choose $Q = 1000$ in our experiment as a trade-off between speed and accuracy.

\begin{figure}[!t]
\centering
\includegraphics[width=\linewidth]{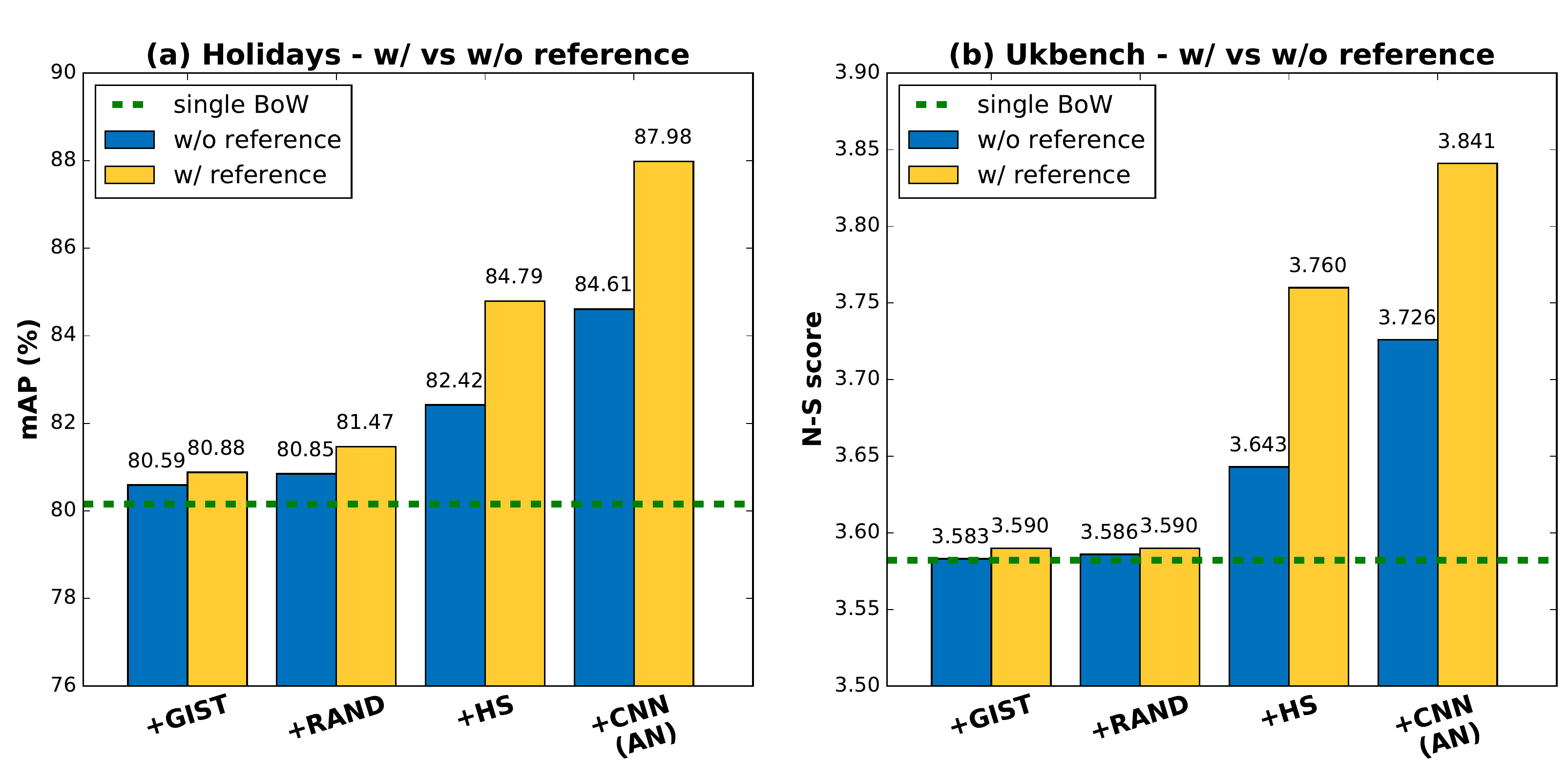}
\caption{The impact of reference. We compare our method with the case where reference is
not used. Four feature combinations are presented, \emph{i.e.}, ``BoW + GIST'', ``BoW + GIST + Random'',
``BoW + GIST + Random + HS'', and ``BoW + GIST + Random + HS + CNN''. The green dash line
represents the BoW results, while blue and yellow bars show results by ``without reference'' 
and ``with reference'', respectively. CNN features are extracted from an ImageNet pre-trained 
AlexNet (AN).}
\label{fig:refvsnoref}
\end{figure}

\textbf{Impact of the reference curve}. To demostrate the effectiveness of reference 
selection, we compare with the case in which no reference is used. In other 
words, the score curve directly undergoes min-max normalization, and the resulting area 
is employed for feature weight estimation. The results are shown in \figurename \ref{fig:refvsnoref}. 
It is clear that, the usage of reference brings benefit for various feature combinations. 
On Holidays and Ukbench datasets, when all five features are fused, the usage of references 
brings improvement of $+3.37\%$ in mAP, and $+0.115$ in N-S, respectively.

\textbf{Comparison with global parameter tuning}. For each features, we assign to it a global weight 
$w^{(i)}, i = 1,...,K$ and search the optimal combination. When fusing five features, we use a step of 0.1 for manual 
tuning. The results are shown in Table \ref{tab:imageretrieval} (Global). We observe that our results are 
comparable, and sometimes even superior to the results of global tuning on the Holidays dataset. 
On Ukbench QAF (ours) also performs comparable to global tuning \textit{without} tuning parameters. It demonstrates the effectiveness of our method. 

Comparing with global parameter tuning, our method is advantageous in two aspects. First, for global tuning, the parameters should be tuned for different datasets. But our method automatically determines the weights, making it more generalizable to different datasets. Second, our method is very efficient in computing the query-adaptive weights, but global feature tuning needs to search over a large parameter space, and this is very time consuming. 



\textbf{Robustness to many ``bad'' features}. When a large number of bad features are present, it is desirable that 
fusion result not be influenced too much. In our experiment, 20 random projection matrices are generated, so 
that we are provided with 20 random projection features. 

\begin{figure}[!t]
\centering
\includegraphics[width=\linewidth]{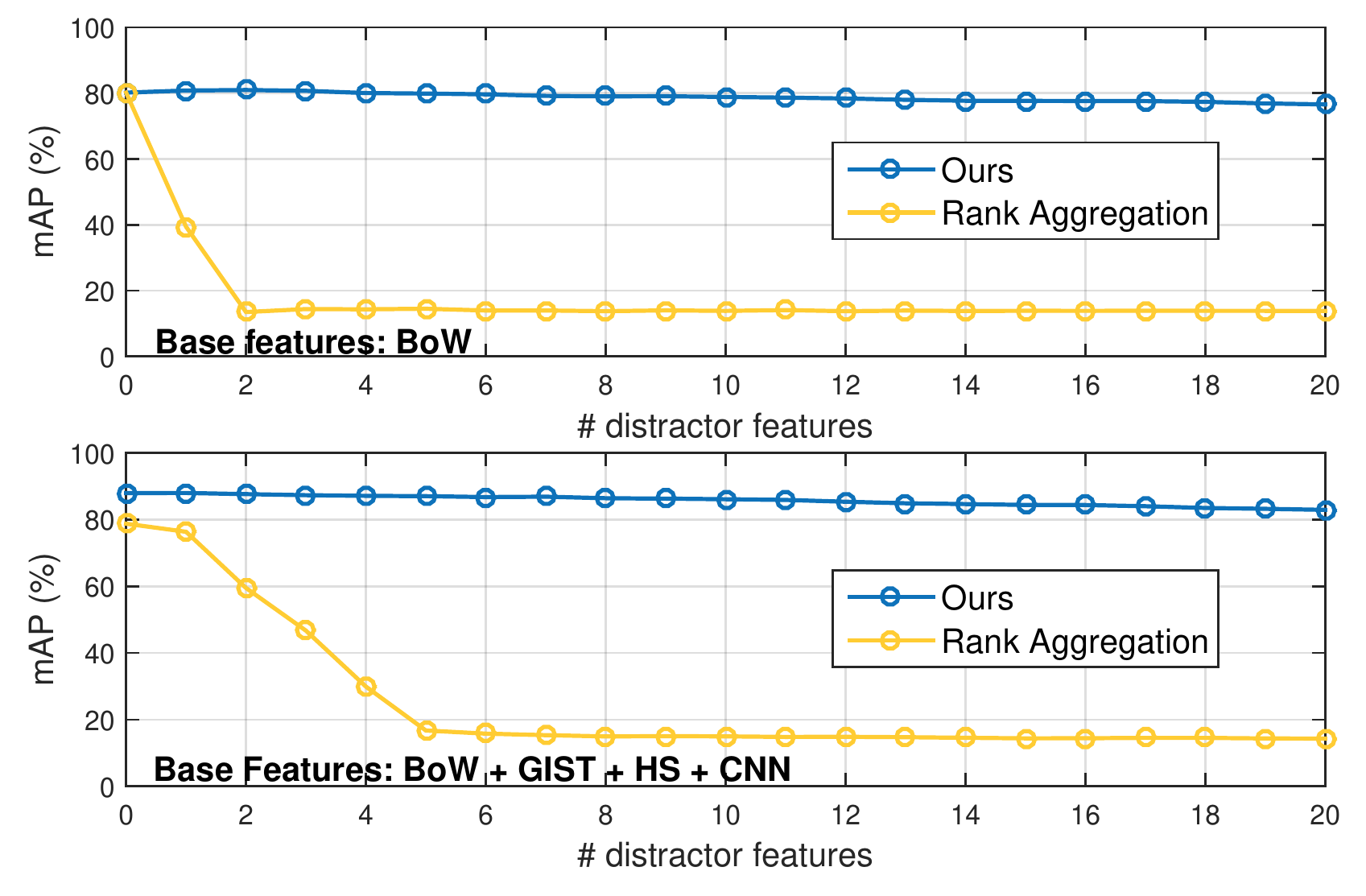}
\caption{Impact of bad features on Holidays dataset. We plot mAP against a increasing number of random features. 
Top: random features are fused with BoW. Bottom: BoW + GIST + HS + CNN (pre-trained AlexNet) is used as baseline. We compare with 
Rank Aggregation \cite{ra}.}
\label{fig:RA}
\end{figure}

We evaluate this property on the Holidays dataset in Fig. \ref{fig:RA}. We compare our method with Rank 
Aggregation (RA). In RA, we compute the median rank of each candidate image over all rank lists obtained by 
different features. We can see that when the number of random features increases, mAP of our method drops 
very slowly, but that of RA decreases dramatically. When as many as 20 ``bad'' features are used, mAP of our 
method drops from $80.16\%$ to $76.58\%$, and from $87.98\%$ to $82.91\%$ for the two base-feature settings, 
respectively. In comparison, RA yields an mAP of $13.85\%$ and $14.29\%$, respectively. Therefore, our 
method is very robust to ``bad'' features.

\begin{figure}[!t]
\centering
\subfloat{\includegraphics[width=0.5\linewidth]{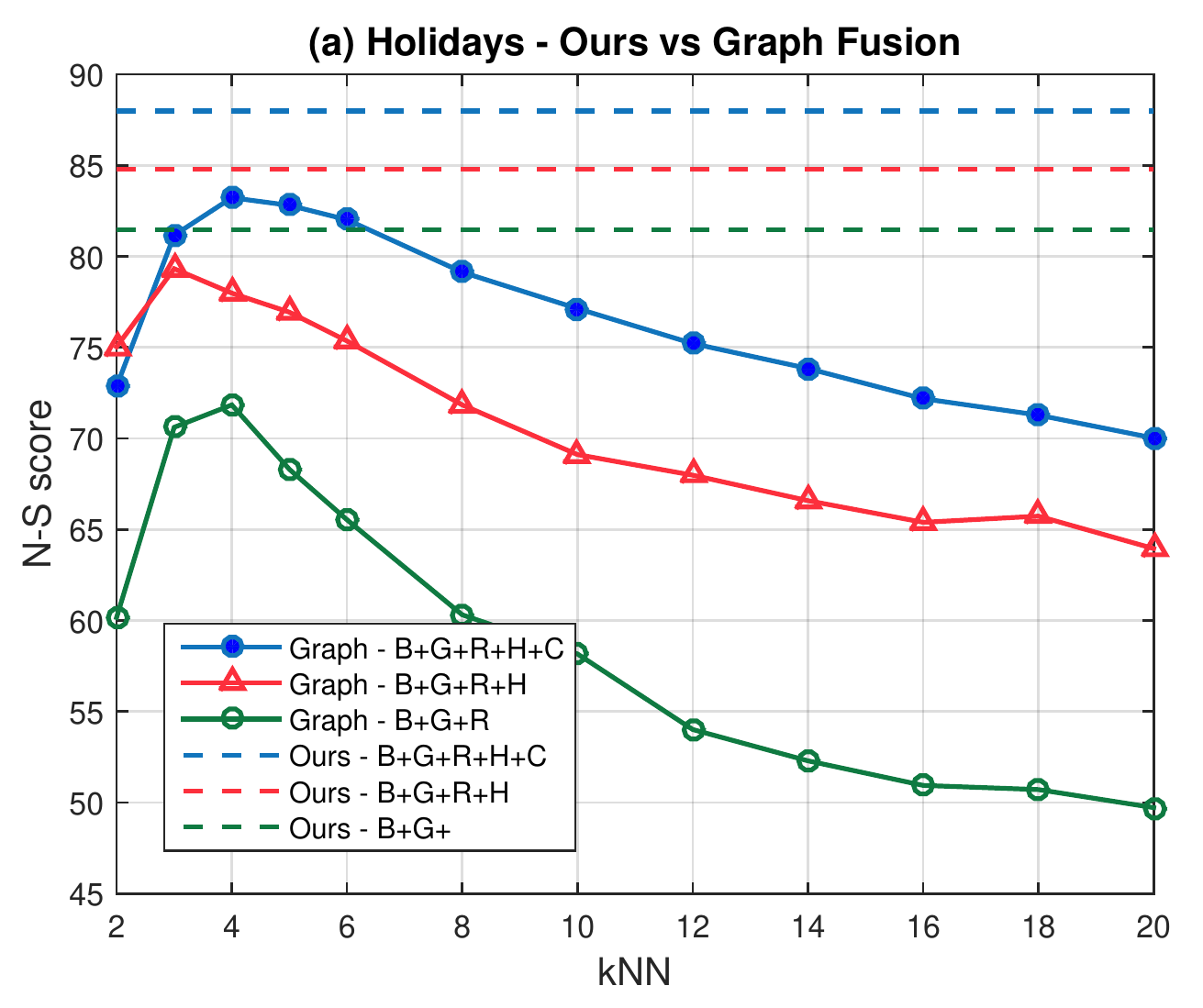}}
\subfloat{\includegraphics[width=0.5\linewidth]{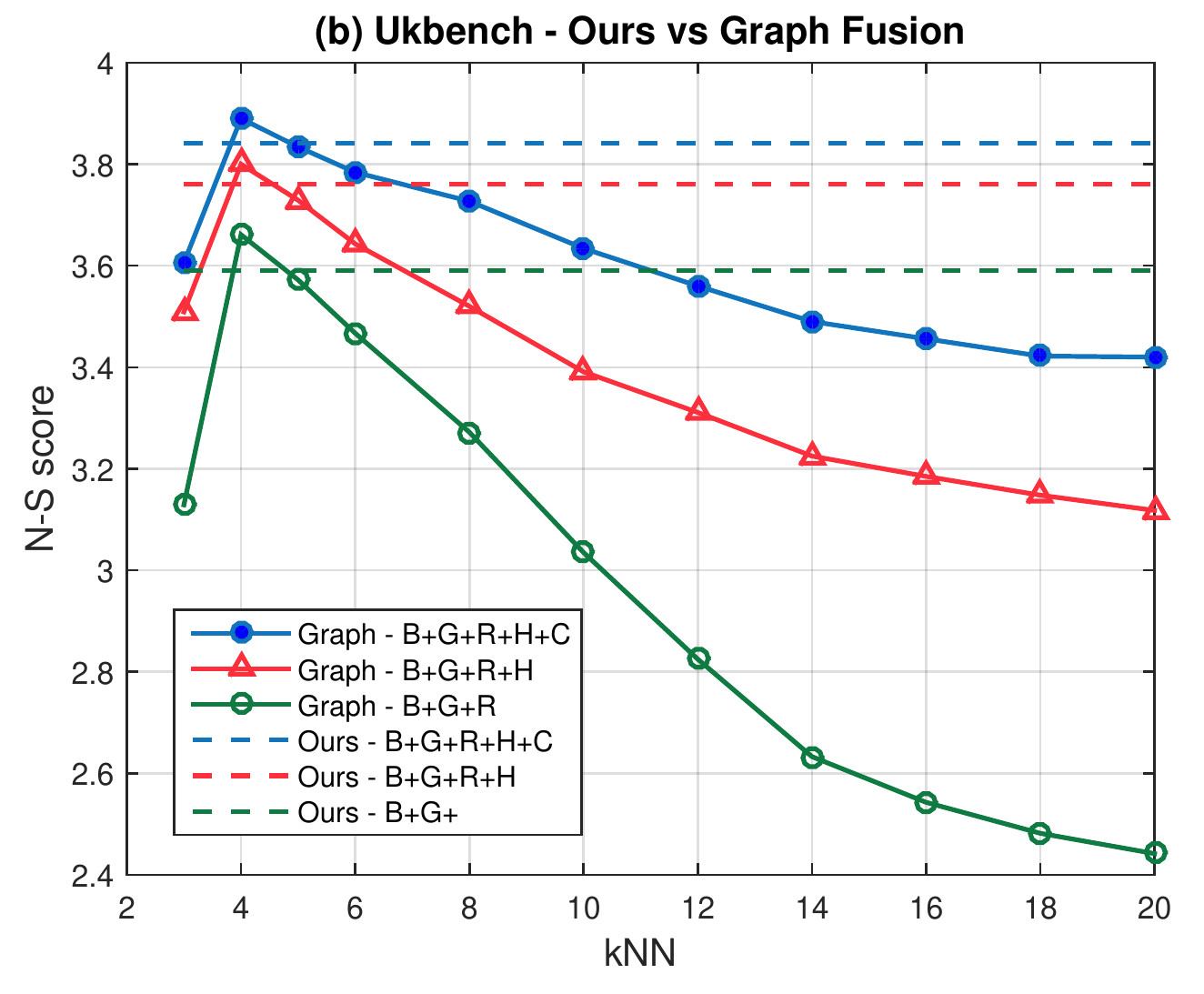}}
\caption{Comparison with graph fusion.On (a) Holidays and (b)Ukbench datasets, three feature combinations
are tested. Abbreviations “B”, “G”, “R”, “H”, and “C” represent BoW, GIST, Random, HSV, and CNN, respectively.
CNN features are extracted by Deep Retrieval model \cite{deepretrieval}. Dashed lines are the results of our
method.“kNN” refers to the key parameter in graph fusion.}
\label{fig:graphfusion}
\end{figure}

\textbf{Comparison with other fusion schemes}. In order to further verify the strength of our method, results 
of two  fusion schemes, \emph{i.e.}, Rank Aggregation~\cite{ra} and Graph Fusion~\cite{graphfusion} are present in 
Table \ref{tab:imageretrieval} and Fig. \ref{fig:graphfusion}. 

On both datasets, Rank Aggregation yields inferior performance to Graph Fusion and our method. 
This is mainly because rank aggregation requires voting from multiple rank results, so that 
when features are complementary to each other, the intersection of the top-ranked candidates 
predicted by different features tends to be small. We can observe from Table\ref{tab:individual}
that when we combine CNN features with BoW and HSV features Rank Aggregation produces good 
results, because these features generate similar top candidates. For comparison, the performance 
drops dramatically when complementary features (GIST, Rand) are fused. 

For graph fusion ,we use the code released by \cite{graphfusion}. Except for the $k$NN value 
(different from the $k$ in our method), we use the default parameters. Results in Fig. 9
indicate that graph fusion is sensitive to parameter $k$NN, which, in order to obtain fine accuracy, 
should be consistent with the average number of true matches in the dataset. For comparison convenience, 
we plot the corresponding results of our method as the dashed lines.

On Holidays dataset, for each feature combination, our method outperforms Graph Fusion. On Ukbench, 
our result is lower than graph fusion only when $k$NN = 4, which is the ideal parameter setting on Ukbench, 
since the number of relevant targets is a constant, \ie, 4, for every query.
Nevertheless, when $k$NN is set to other values, the performance of graph fusion drops. Moreover, when “bad”
features, such as GIST and Random are used, graph fusion does not have a “fall back” mechanism (in fact, it treats all the features as equally important). Considering that our method is robust to parameters
(see \figurename \ref{fig:param:uv} and \figurename \ref{fig:param:q}), we speculate that our method yields more stable and accurate performance than graph fusion.

\textbf{Complexity and Scalability}. The proposed method can be broken down into two phases, \emph{i.e.}, offline reference 
codebook generation and online query adaptive weight estimation. For the offline procedure, the
complexity is $O(Q|\mathcal{I}|)$ where $Q$ is the codebook size and $\mathcal{I}$ is the 
irrelevant dataset for reference construction. Note that for a specific feature, we only need to construct
the reference codebook once for all, so it does not matter if $|\mathcal{I}|$ is slightly large. 

In experiment, we perform a large-scale experiment by combining MirFlickr1M dataset \cite{1mdistractor} 
with Holidays dataset to test the scalability of the system. As noted in Section \ref{sec:unsupervised}, we down-sample the initial score
lists and references to a length of 1000. Moreover, dimension of all four global features are
reduced to 128-D by PCA. Our experiments are performed on a server with 3.46 GHz CPU and 128 GB memory. 
CNN features are extracted with a GTX 780 Ti GPU. The average query time and the breakdown are shown in Table \ref{tab:breakdown}. 
Our method adds little extra time in inference selection. Moreover, the storage of the reference
codebook costs only 7.63MB extra memory. Therefore, our method is efficient in terms of memory and
time cost. As comparison, graph fusion also adds neglectable time cost in graph analysis. However, 
it needs to compute pairwise distances of the union set of query and database images, which is 
$O((|\mathcal{Q}|+|\mathcal{D}|)^2)$ in complexity, where $\mathcal{Q}$ and $\mathcal{D}$ are the 
query set and database set respectively. When the database grows larger, the time cost
increases dramatically. There is no such procedure in our method, so we only need to compute 
pair-wise distance between query and database which is only $O(|\mathcal{Q}||\mathcal{D}|)$.

\begin{table}[t]
\caption{Average query time (s) of different steps on Holidays + 1M, feature extraction and quantization time excluded.}
    \centering
    \begin{tabular}{|l||ccc|}
    \hline
        Stage               &     BoW &   Glob. Feat. &  Ref. Selection  \\
    \hline
        Avg. Time (s)        &  1.95  &  0.96& 0.01\\
    \hline
    \end{tabular}
    
    \label{tab:breakdown}
\end{table}

\subsection{Results on Person Recognition}
\label{sec:exp:personrecognition}
\begin{figure*}
    \centering
    \subfloat{\includegraphics[width=0.48\linewidth]{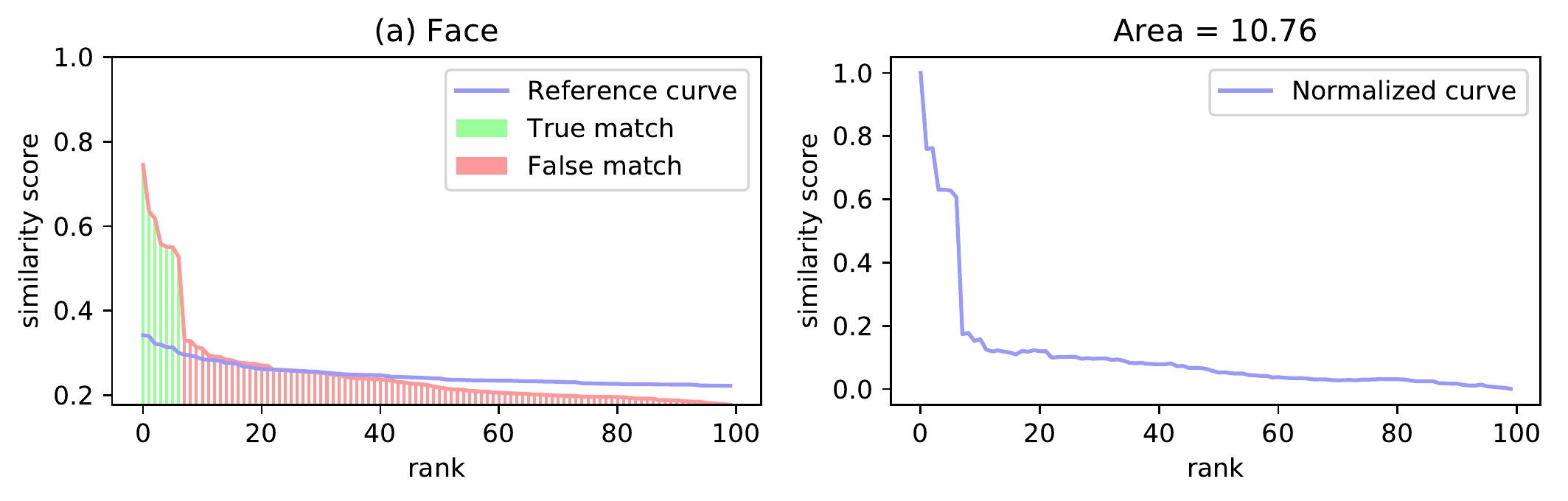}}
    \hfill
    \subfloat{\includegraphics[width=0.48\linewidth]{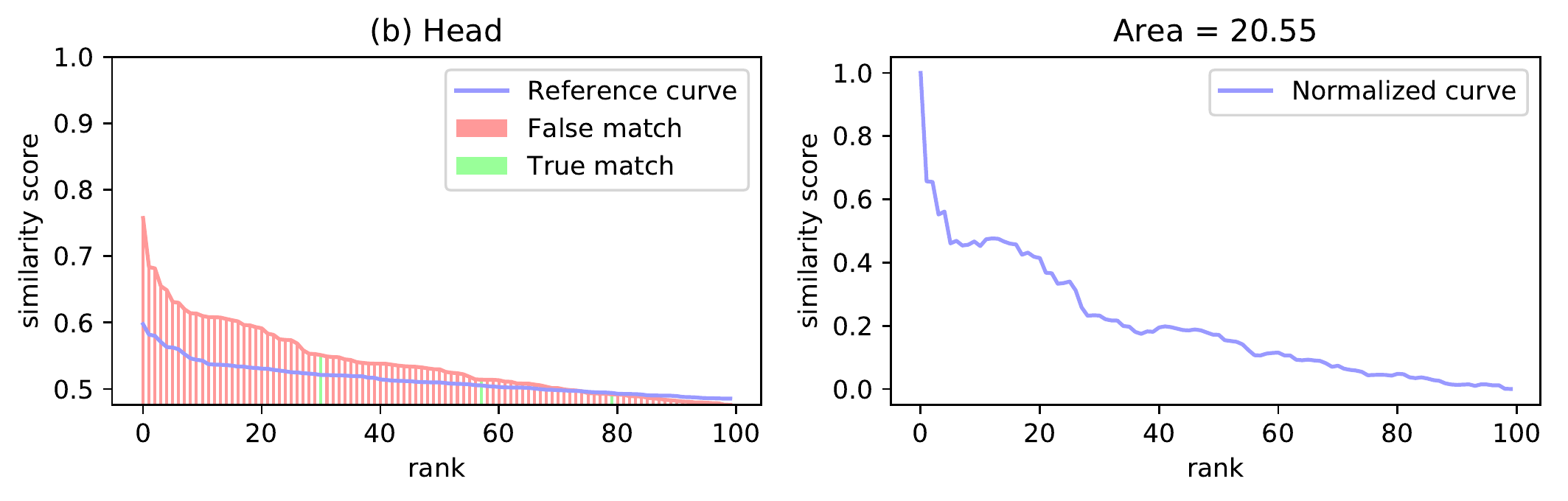}}
    \par
    \vspace{-0.15in}
    \subfloat{\includegraphics[width=0.48\linewidth]{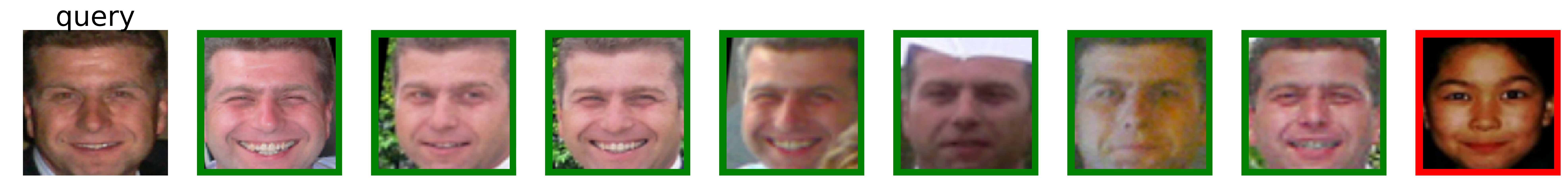}}
    \hfill
    \subfloat{\includegraphics[width=0.48\linewidth]{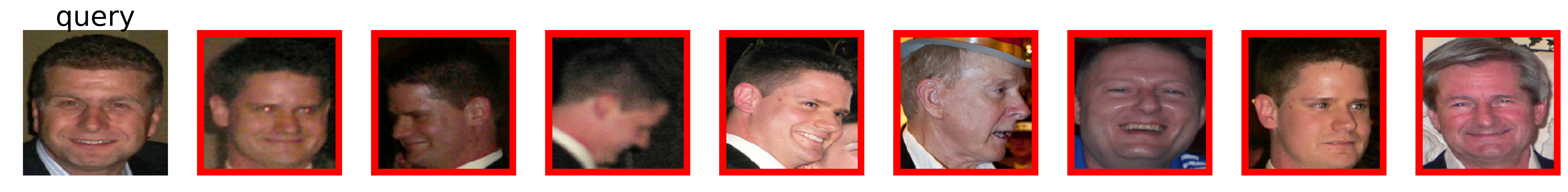}}
    \par

    \subfloat{\includegraphics[width=0.48\linewidth]{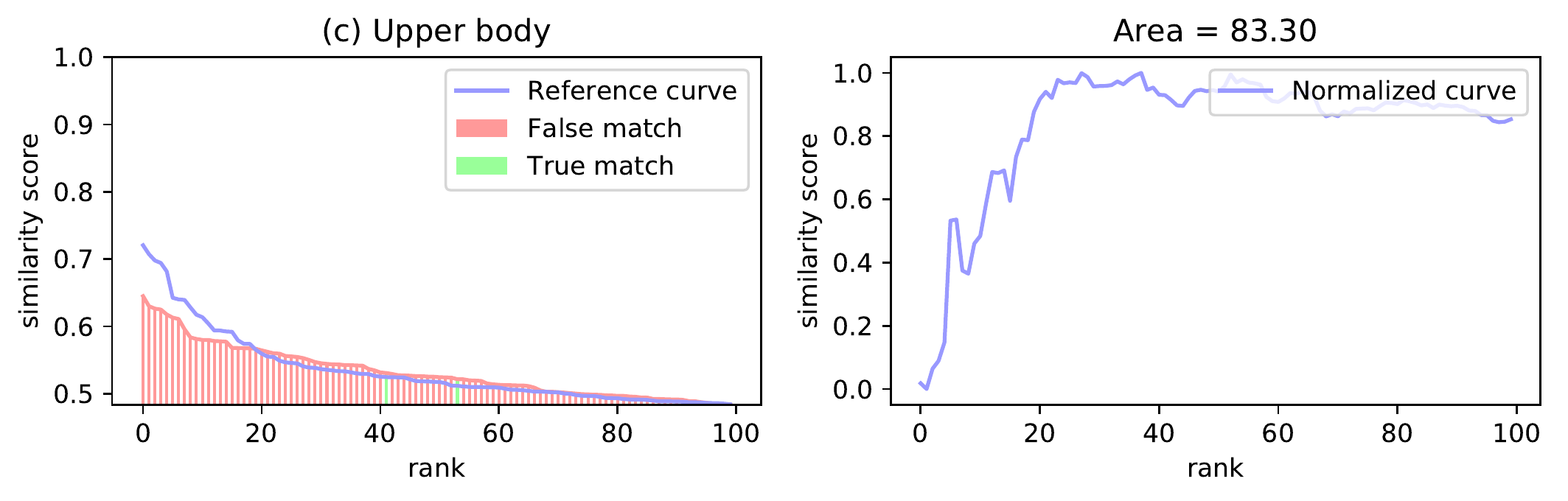}}
    \hfill
    \subfloat{\includegraphics[width=0.48\linewidth]{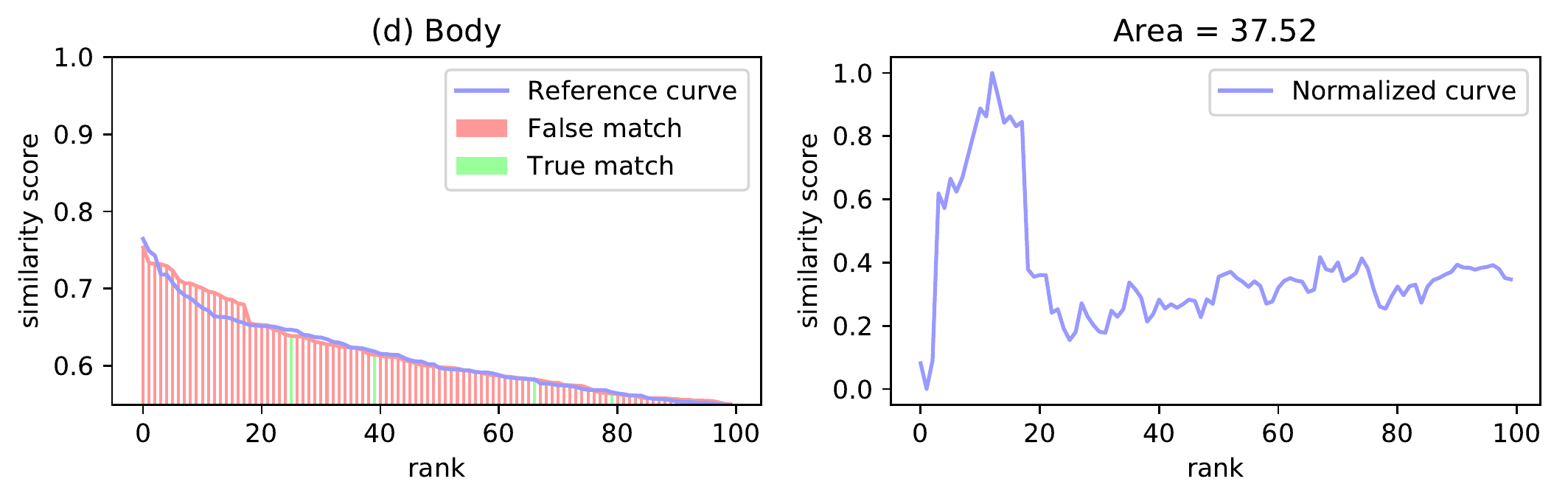}}
    \par
    \vspace{-0.15in}
    \subfloat{\includegraphics[width=0.48\linewidth]{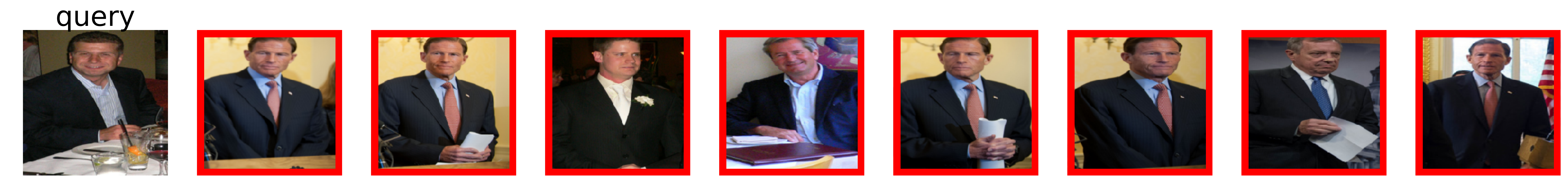}}
    \hfill
    \subfloat{\includegraphics[width=0.48\linewidth]{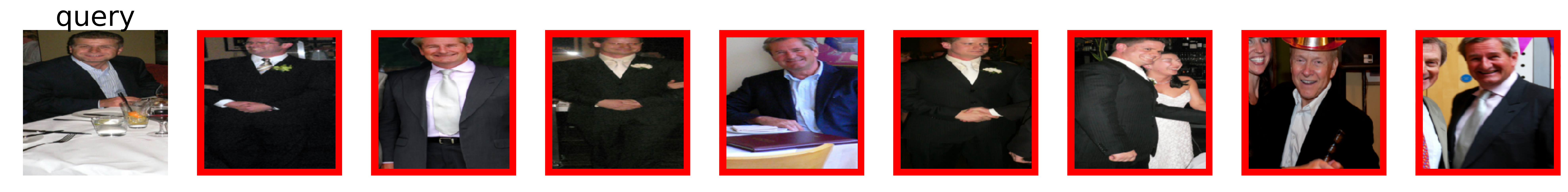}}
    \par

    \subfloat{\includegraphics[width=0.33\linewidth]{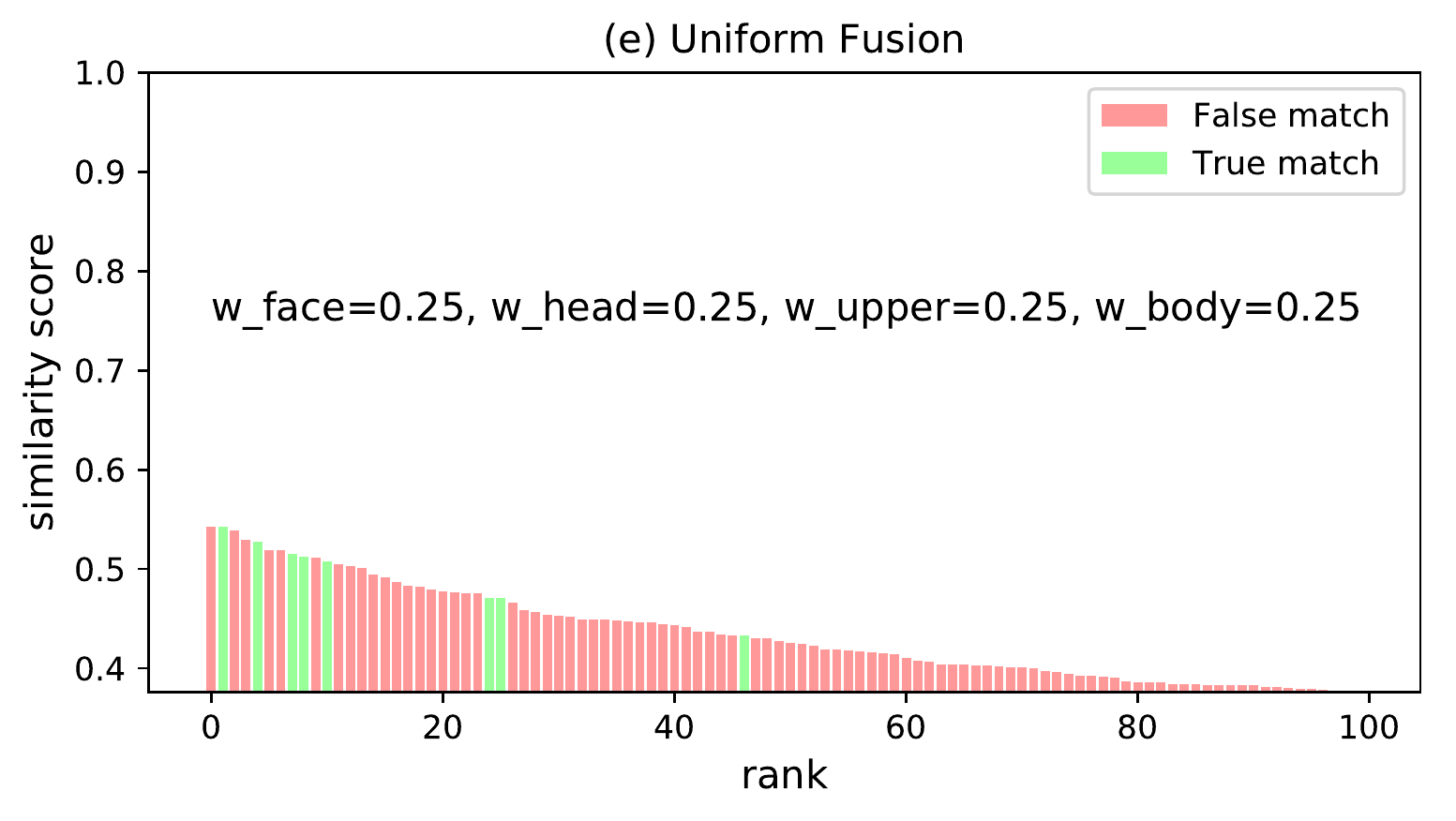}}
    \hfill
    \subfloat{\includegraphics[width=0.33\linewidth]{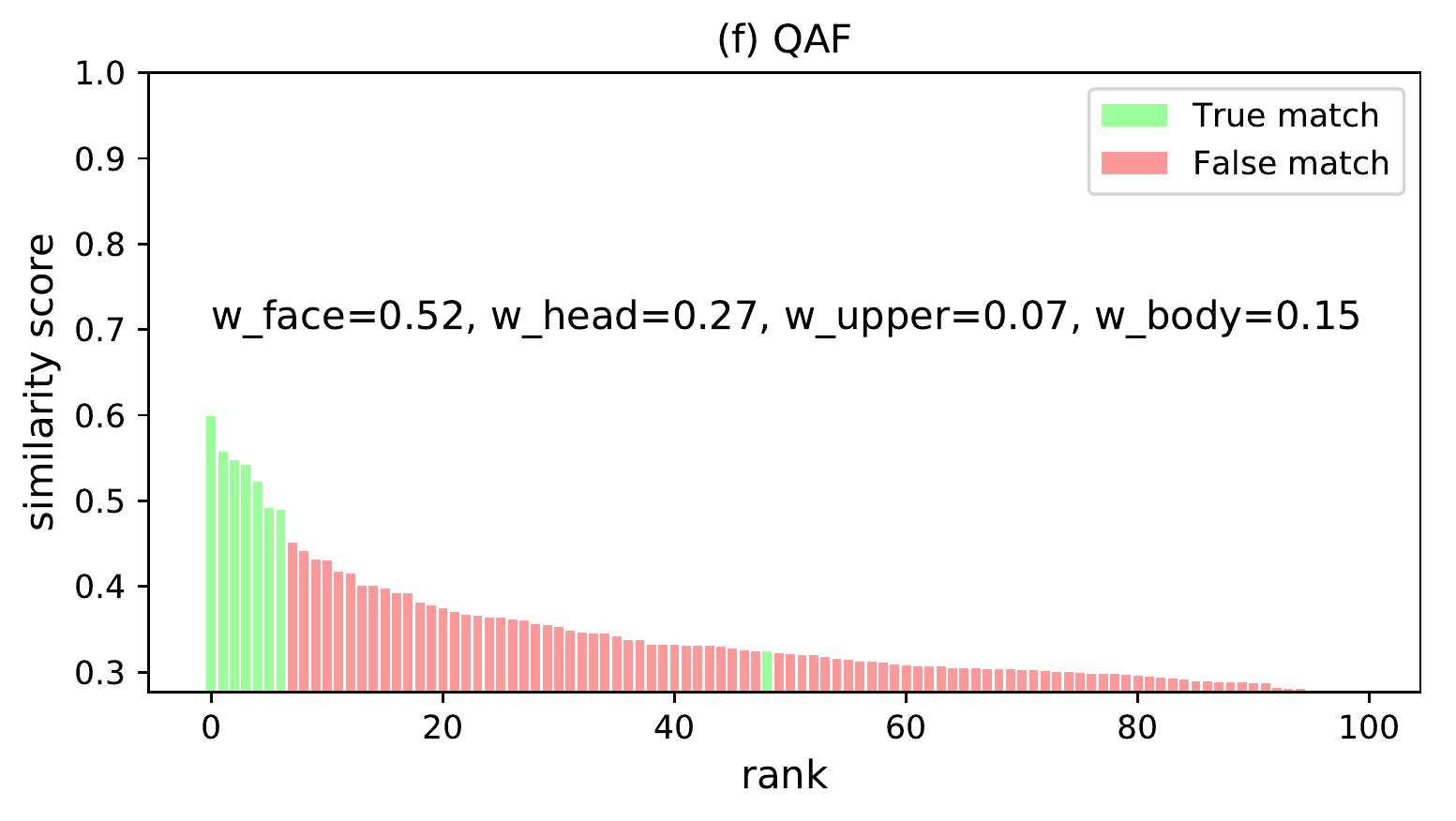}}
    \hfill
    \subfloat{\includegraphics[width=0.33\linewidth]{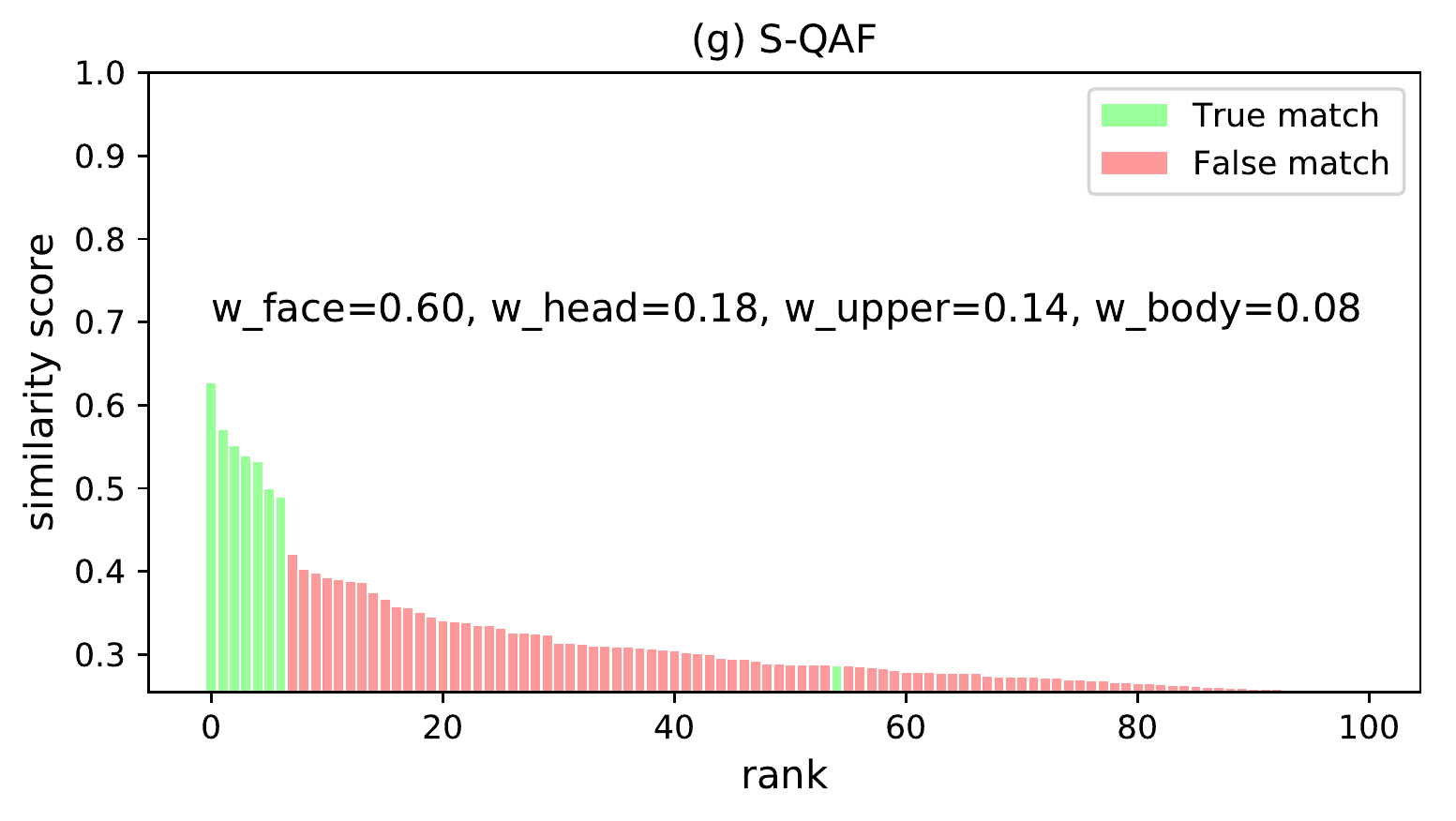}}
    \caption{Illustration of query adaptive fusion for person recognition. (a)-(d) present the score curves of face, 
    head, upper body and holistic body features, the reference curve and normalized curve (for
    unsupervised query adaptive fusion), and also the top 8 candidate of the query for each feature. (e) presents the fusion result
    of uniform fusion, \emph{i.e.}, assigning all features with the same importance. (f) and (g) shows the fusion results 
    of the unsupervised query adaptive fusion (QAF) and the supervised query adaptive fusion (S-QAF), respectively. 
    For this query, we observe that the face feature is the most important. Both QAF and S-QAF succeed in estimating 
    the feature importance, while S-QAF is slightly superior since it assigns a larger weight to the face feature. }
    \label{fig:pipa}

\end{figure*}

\textbf{Dataset and Evaluation Metric}. We use the People in Photo Album (PIPA) dataset~\cite{pipa} for both training and
evaluation. To our knowledge, the PIPA dataset is the first dataset collected for unconstrained person recognition. 
It consists of over 60,000 instances of ~2,000 individuals collected from public Flickr photo albums. For 
each instance, a head bounding box is provided, regardless if the instance is in frontal view or in back view. For 
those in back view, annotators infer their identity by the clothes, the surroundings, and also the relationship 
with other identities, which guarantees the accuracy of the annotations. PIPA is challenging when using single visual cues due 
to the occlusion with other people, viewpoint, pose and variations in clothes. For instance, if we only pick faces
as the retrieval cue, we would see that only $52\%$ of the people have high resolution frontal faces suitable for 
recognition. Clothes are also a good appearance feature, but difficulties arise when different persons wear similar clothes (which is common in occasions like festivals or celebrations), or the same person appears in multiple albums wearing different 
clothes. All of the above difficulties make it very suitable to apply query adaptive fusion to person recognition. 

The original test setting of PIPA in \cite{pipa} considers person recognition as a classification task. In \cite{pipa},
a feature extractor is first trained on the training set, and then a classifier is trained on a subset of the 
testing set that contains all the identities of the testing set. This setting actually might deviate from practice, because it 
is impractical to train a new classifier every time a new identity is introduced during testing. 

In this paper, we formulate person recognition as an open-set retrieval problem, where we use the top-1 accuracy as the evaluation metric. The top-1 accuracy is
comparable to the recognition accuracy in the original close-set classification setting, since it can be regarded 
as the recognition accuracy of a nearest neighbor classifier. The open-set setting is closer to practice because it does not need retrain classifiers when new testing persons are added. 

In our protocol, the testing set is split into a query set and a gallery set. Following the setup in \cite{pr1}, we split the test set in four different ways, namely \textit{original}, \textit{album}, \textit{time}, and \textit{day}. The \textit{original} 
setup is the easiest one, in which a query instance may have very similar appearance with its true matches in the gallery. 
The other three are more challenging. For example, the \textit{day} setup enforces the query instances has appearance changes 
compared to the true matches, while the \textit{time} setup emphasizes the temporal distance between query set and the gallery set. We test all the four setups. For each setup, we switch the query set and gallery set, so that we can have two testing results. The two results are averaged to obtain the final performance. 

\textbf{Implementation Details}. According to Section \ref{sec:person_recognition}, for each of the four body parts, namely face, head, upper body and holistic body, we separately train a CNN feature extractor. The face feature extractor is a 20-layer ResNet~\cite{resnet} trained with CASIA-WebFace dataset
\cite{casia} and the A-Softmax loss \cite{sphereface}. All faces are detected by MTCNN\cite{mtcnn} and aligned with five facial
landmarks. We use the Inception-v3 \cite{inceptionv3} architecture for the other three feature extractors, and train them on the PIPA
training set with the softmax loss. Hyperparameters for training face feature extractors are 
the same as in \cite{sphereface}. For all the extractors, we use SGD as the optimizer with learning rate 0.1, and decay by 0.1 
at the $30^{th}$, the $45^{th}$ and the $53^{rd}$ epoch. The total number of epochs is set to 60, and the batch size is set to 128.

\noindent
\textbf{Performance of Individual Features.}
\begin{table}[t]
    \caption{Top-1 accuracy of individual features on the PIPA dataset.} 
    \centering
    \begin{tabular}{l|cccc}
    \toprule
         Feature &  Original & Album     &   Time    & Day   \\
    \hline
         Face&      65.08   &   63.49   &   61.04   &  61.22\\ 
    \hline
         Head&      76.07   &   66.97   &   56.73   &  35.60\\
    \hline
         Upper Body&75.04   &   64.14   &   51.68   &  22.80\\
    \hline
         Body&      68.11   &   57.01   &   43.16   &  16.93\\
    \bottomrule
    \end{tabular}

    \label{tab:pr:individual}
\end{table}
Table \ref{tab:pr:individual} showcases the performance of individual features in each setup. In the \textit{original}
setup, the head feature and upper body feature present good retrieval accuracy. This is expected, because the head and the 
upper clothes are two primary cues to search the same person. Theoretically, the holistic body should be more 
informative than the upper body, but in practice it suffers from misalignment, for example,
if a person is sitting or lying, the holistic body bounding box would be inaccurate. Another observation is that the 
face feature has stable performance when the testing setup changes. In the \textit{Day} setup, the face feature remains 
an accuracy of $61.22\%$, while performance of other features drops dramatically, especially the upper body and the 
holistic body features. This is because in the \textit{Day} setup identities wear different clothes between the
query set and the gallery set, so that the face feature and head feature are more reliable. 

\begin{table}[t]
    \caption{Comparison with state-of-the-art results and other fusion schemes on PIPA.}
    \centering
    \begin{tabular}{l|cccc}
    \toprule
        Method & Original & Album & Time & Day \\
    \hline
        PIPER~\cite{pipa} &83.05&-&-&- \\
        naeil~\cite{pr1}&86.78&78.72&69.29&46.61 \\
        Pose-aware~\cite{pr2}&89.05&83.27&74.84&56.73 \\
        COCO~\cite{cocoloss}&92.78&83.53&77.68&61.73 \\
        RANet~\cite{ranet}&89.73&85.33&80.42&67.16 \\
    \hline
    \hline
        $Global^{original}$& 88.52&83.52&77.08&62.61\\
        $Global^{album}$&87.74&83.68&77.91&67.60 \\
        $Global^{time}$&87.74&83.68&77.91&67.60 \\
        $Global^{day}$&85.59&82.34&77.09&68.85 \\
    \hline
    \hline
        Uniform&88.26&83.40&77.13&64.60\\
        QAF&88.01&84.15&78.33&66.41\\
        S-QAF&89.79&85.21&79.45&68.20\\
    \bottomrule
        
    \end{tabular}

    \label{tab:pipa}
\end{table}

\textbf{Unsupervised v.s. Supervised Query Adaptive Fusion}. In this section, we compare the proposed unsupervised query 
adaptive fusion scheme (QAF) with the supervised query adaptive fusion scheme (S-QAF). For QAF, we set $k$NN = 5, 
$u= 10$ , $v = 400$, and $Q = 1000$. The reference codebook is constructed with the PIPA training set. In the construction of each reference codebook, 
we first select one image of an identity as the query and use the images of other identities as database images to make sure there 
is no true match. For S-QAF, we 
train the S-QAF module with the PIPA validation set. Qualitative and quantitive comparisons are present in Fig. \ref{fig:pipa} 
and Table \ref{tab:pipa} respectively. 

Figure \ref{fig:pipa} shows a typical query instance in the \textit{Day} 
setup: the database instances of the same identity wear different clothes \emph{w.r.t} the query. In this case, the upper body and holistic 
body features are unreliable, while the face feature is discriminative. As shown in Fig. \ref{fig:pipa} (f) and (g), both 
QAF and S-QAF succeed in estimating the importance of the features. Moreover, S-QAF seems superior because it assigns a 
higher weight to the face feature. Table \ref{tab:pipa} compares QAF and S-QAF with the Uniform Fusion baseline in terms of 
retrieval accuracy. In the Uniform Fusion baseline, all the features are fused with the same fixed weight. Both QAF and S-QAF 
outperforms the baseline, indicating that the fusion schemes are effective. It is expected  that S-QAF performs better than QAF, 
because the weights are learned end-to-end with a loss function which enlarges the gap between true matches and false matches.

\begin{figure}
    \centering
    \subfloat{\includegraphics[width=0.5\linewidth]{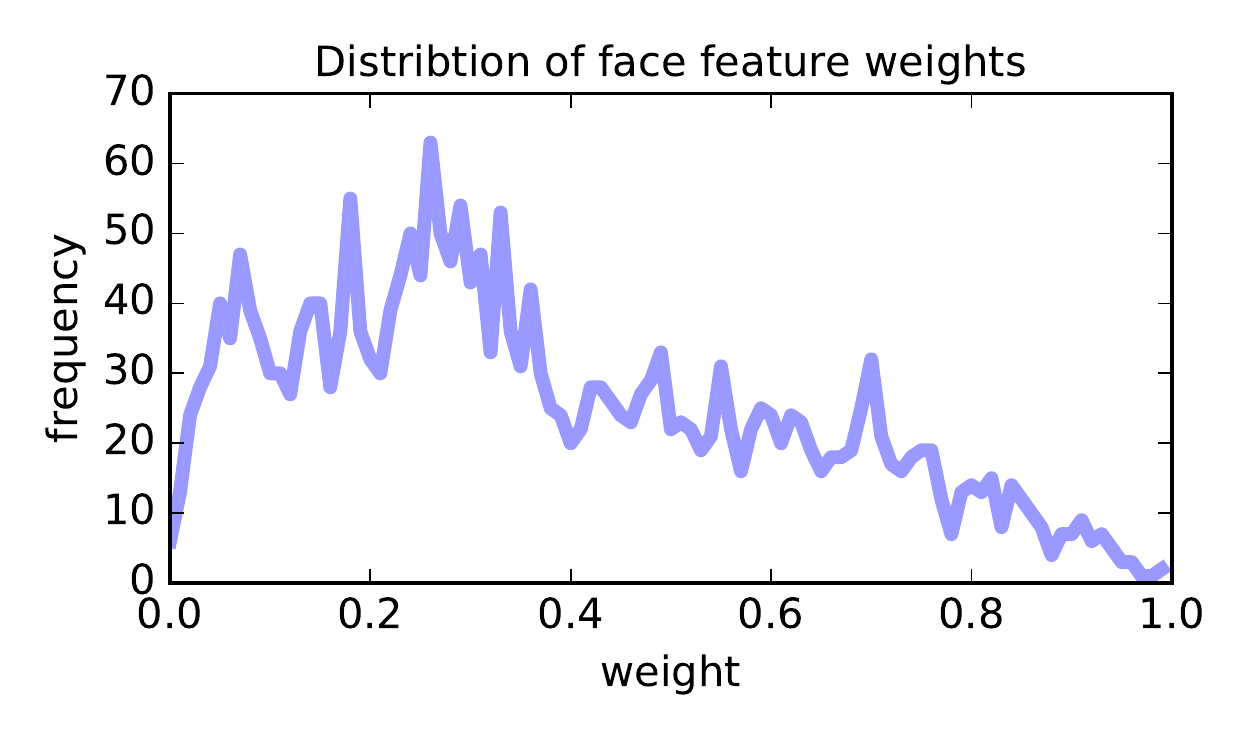}}
    \subfloat{\includegraphics[width=0.5\linewidth]{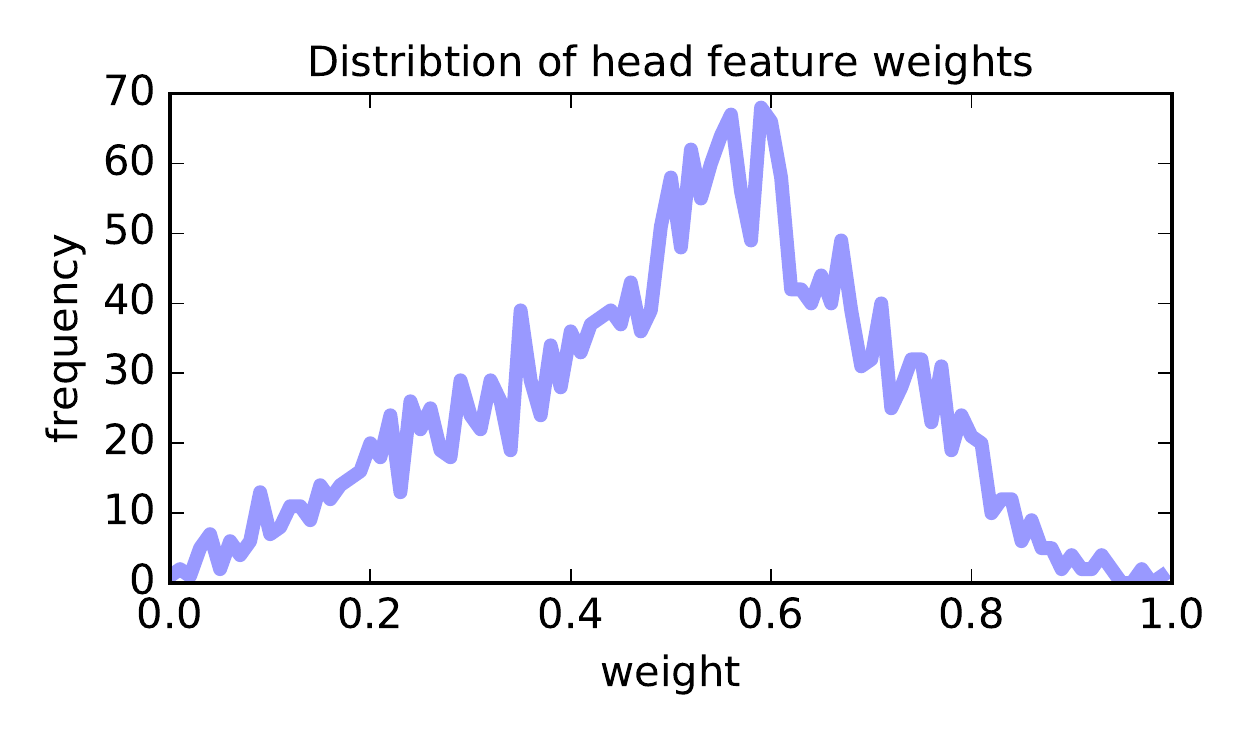}}
    \par
    
    \subfloat{\includegraphics[width=0.5\linewidth]{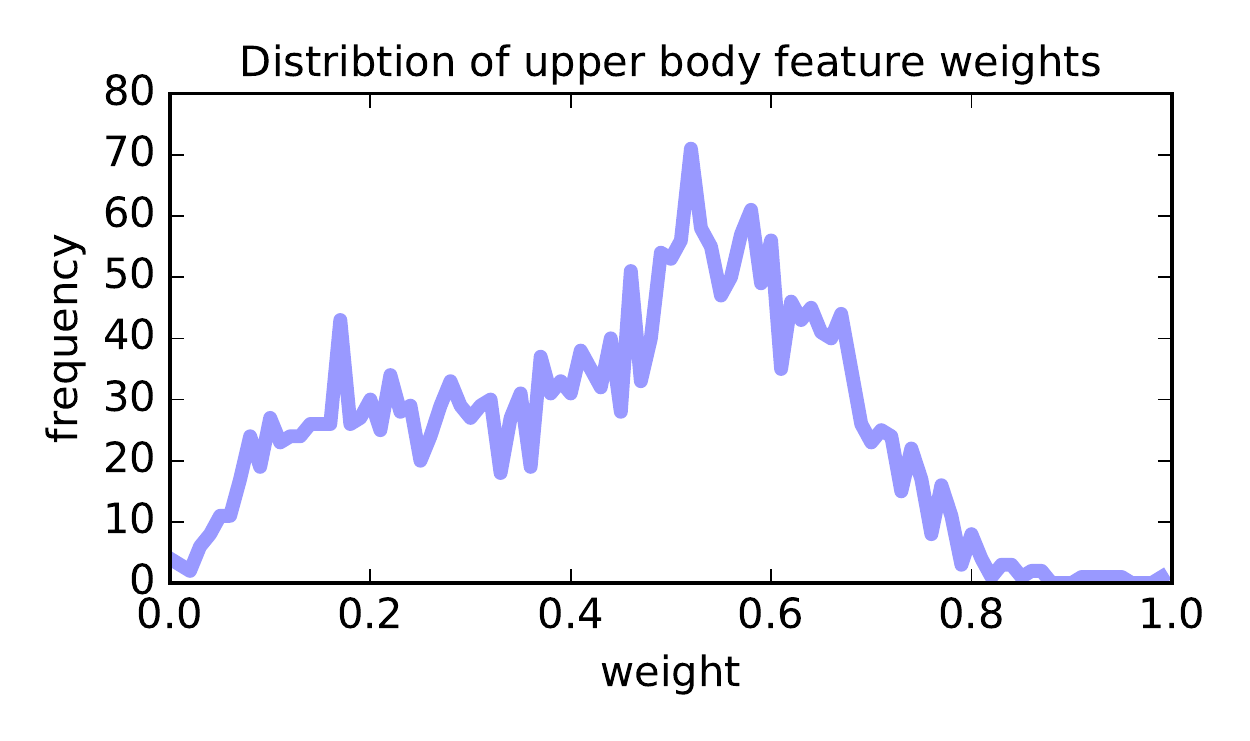}}
    \subfloat{\includegraphics[width=0.5\linewidth]{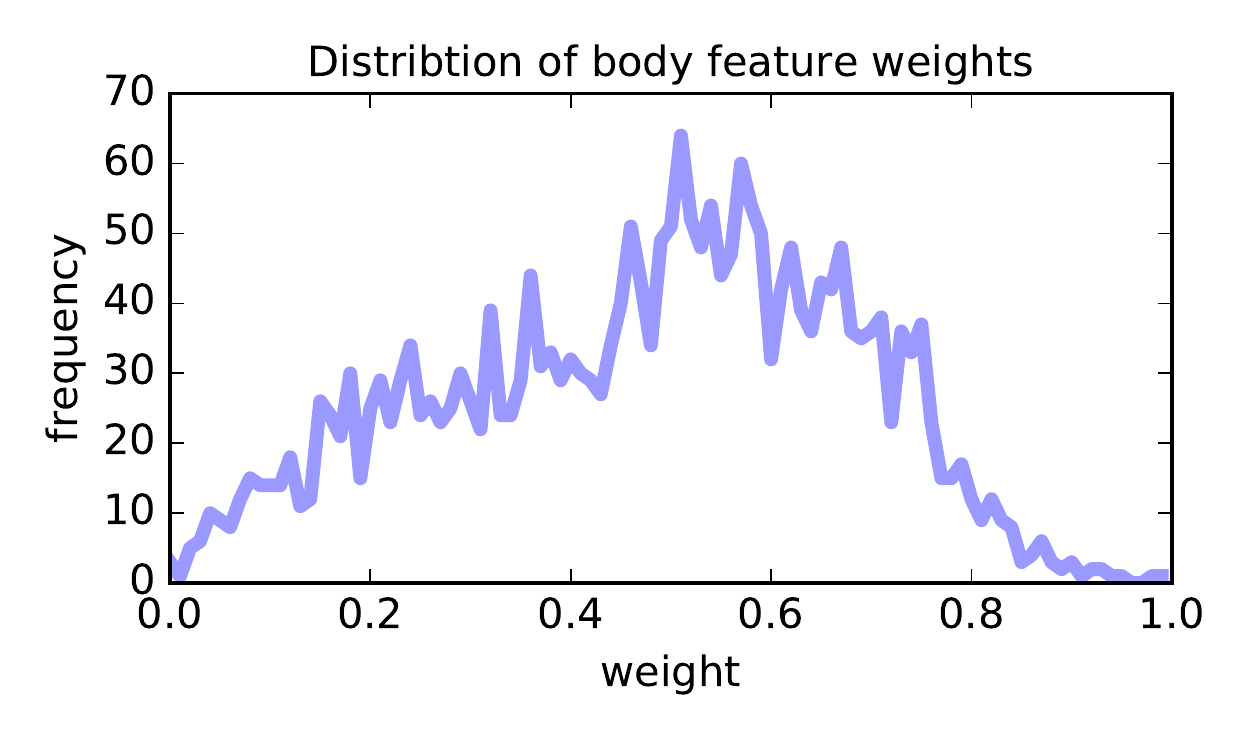}}
    \caption{Distribution of the feature weights of face, head, upper body, and holistic body generated by S-QAF under the PIPA \textit{Day} setup.}
    \label{fig:stat}
\end{figure}

\textbf{Comparison with global fusion}. In order to validate the effectiveness of QAF and S-QAF, we perform grid search to seek 
optimal global weights for each testing setups. By ``global'', we assign a constant weight to a component feature, \emph{i.e.}, not ``query adaptive''. The step of the grid search is set to $0.1$. The optimal results are demonstrated in the second 
part in Table \ref{tab:pipa}. As expected, the optimal global weights for different testing setups are different. For example, 
the optimal weights are $(0.2,0.4,0.2,0.2)$ and $(0.4,0.3,0.2,0.1)$ for the \textit{Original} and the \textit{Day} setups, respectively. QAF exceeds the global fusion 
in two test setups (\textit{Album} and \textit{Time}), and S-QAF exceeds global fusion in three setups (\textit{Original},\textit{Album} 
and \textit{Time}). For those setups where QAF and S-QAF do not outperform the global fusion, the results are also competitive. The 
main benefit of QAF and S-QAF is their robustness to the change of test set. In other word, QAF and S-QAF performs well when the appearance change between query and gallery is large (the \textit{Day} setup) and when the appearance change is small 
(the \textit{Original} setup). In fact, if we assign global weighs to the heterogeneous features, we cannot find a single set of optimal weights 
that suits all test setups. It means we need to search for the weights for each datasets, which is very time-consuming. We also visualize the distribution of weights generated by S-QAF on PIPA \textit{Day} test split.  
The results are shown in Fig. \ref{fig:stat}. The feature Weights are distributed scatteredly in the histogram. This indicates that the feature weights are determined in a query-adaptive manner and that the weights differ from query to query. 

\textbf{Comparison with state of the art}. Comparison with the state-of-the-art results on PIPA are showcased in the Table 
\ref{tab:pipa}. In some of the compared methods, auxiliary information is utilized to improve retrieval accuracy. For example, in \cite{pipa} and \cite{pr2}, the human pose information is integrated.  In \cite{pr1} and \cite{ranet}, the scene information or 
the relationship among persons are analyzed. Liu \etal \cite{cocoloss} use the most similar setting to us, \emph{i.e.}, only the human appearance is 
used, and our performance is comparable with \cite{cocoloss}. Comparing with these methods, we observe from Table \ref{tab:pipa} that the reported accuracy is very competitive.

\textbf{Complexity and scalability}. Tested on a server with 3.46 GHz CPU, 128 GB memory and a Titan Xp GPU, the average 
time cost for feature extraction for one query is 23.5 ms, 1.6 ms and 0.6 ms respectively.  On top of 
the time for feature extraction, QAF and S-QAF add very little extra time cost in the fusion process. As for the memory cost, the codebook size of QAF is 8.46 MB, 
while the size of S-QAF model is only 72.8 KB. Moreover, the computation cost of QAF and S-QAF does not increase with 
the size of the database, so both methods have good scalability.

\subsection{Results on Pedestrian Retrieval}\label{sec:resultsreid}

\begin{figure}
    \centering
    \subfloat{\includegraphics[width=\linewidth]{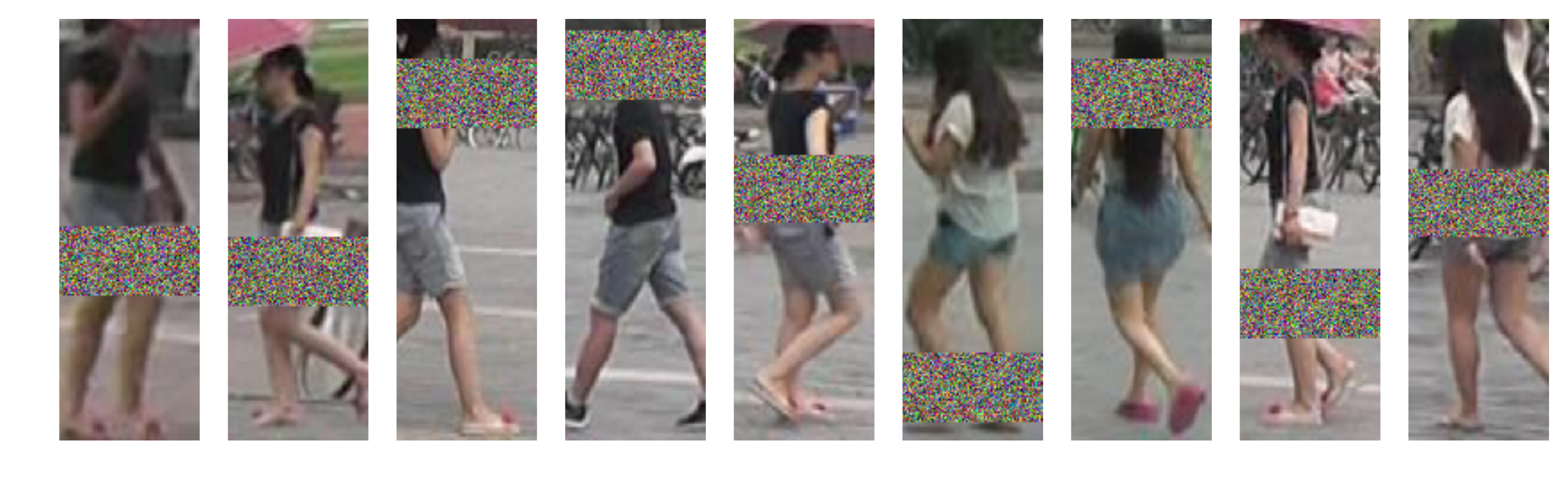}}
    \par
    \subfloat{\includegraphics[width=\linewidth]{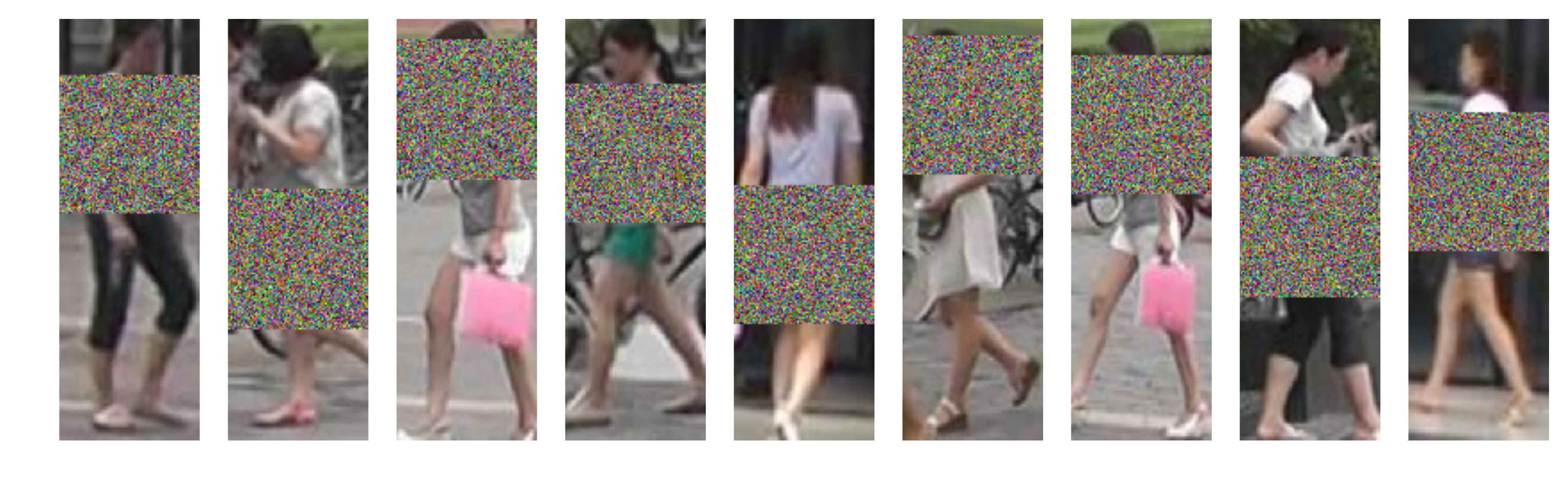}}
    \caption{Illustration of images under occlusion levels of $\sfrac{1}{6}$  (first row) and 
      $\sfrac{1}{3}$ (second row). The definition of occlusion levels can be viewed in Section \ref{sec:resultsreid}. }
    \label{fig:occluded}
\end{figure}
\begin{table}[t]
\label{tab:reid}
\caption{Retrieval accuracy v.s occlusion level. Rank-1 accuracy and mAP are reporeted on the Market-1501 dataset. Two methods are compared, \emph{i.e.,} feature concatenation (Concat.) and score-level averaging (Avg.).}
\centering

   \begin{tabular}{|c|c|p{1.3cm}<{\centering}|p{1.3cm}<{\centering}|p{1.3cm}<{\centering}|}
    \hline
    \multicolumn{2}{|c|}{\multirow{2}{*}{}}&\multicolumn{3}{c|}{mehthod}\\
    \cline{3-5}
    \multicolumn{2}{|c|}{}&Concat. & Avg. & S-QAF\\
    \hline
    \multirow{2}{1.2cm}{No Occ.}&mAP&75.4 & 75.2 & 73.5\\
    \cline{2-5}
    & Rank-1 &91.9 & 91.8&91.0 \\
    \hline
    \multirow{2}{1.2cm}{$\sfrac{1}{6}$ Occ.}&mAP &28.3 & 26.5 & 32.8\\
    \cline{2-5}
    & Rank-1 & 72.3 & 70.9 & 73.5 \\
    \hline
    \multirow{2}{1.2cm}{$\sfrac{1}{3}$ Occ.}&mAP& 12.2 & 11.3& 15.4\\
    \cline{2-5}
    & Rank-1 & 54.7 & 53.1 & 56.4\\
    \hline
    \end{tabular}
\end{table}

\textbf{Dataset and Evaluation Metric}. To further investigate the application scope of the proposed query adaptive fusion, we  perform pedestrian re-identification (retrieval) experiment on the Market-1501 dataset \cite{market}. 
It is one of the most widely used datasets in pedestrian retrieval, which consists of 19,732 gallery images 
and 12,936 training images from 1,501 identities. We use mAP and top-1 accuracy of the Cumulative Matching 
Characteristics (CMC) curve as evaluation metrics. 

\textbf{Occluded Pedestrian Retrieval}. As it is designed to be, query adaptive fusion has the ability to estimate 
the importance of features. To validate this point, we perform experiment on occluded pedestrian retrieval, 
so that the part features have different discriminative ability.  In order to simulate occlusion, we add a rectangle mask generated by Gaussian noise. Two occlusion levels are used,  \emph{i.e.}, $\sfrac{1}{6}$ and $\sfrac{1}{3}$, which denote the ratio of size of the mask to the size of the image. Note that the mask is located randomly. Examples are shown 
in Fig. \ref{fig:occluded}.
We use the part-based convolutional baseline (PCB) \cite{pcb} as the pedestrian descriptor, which outputs six part-based features for a 
pedestrian image. The features are fused by three different methods, \emph{i.e.}, directly concatenating at feature level, 
averaging at score level, and fusion with S-QAF at score level. We then observe how the performance of different fusion methods 
vary \emph{w.r.t} different occlusion levels.  

The comparison results are shown in Table \ref{tab:reid}. Several conclusions can be made. 
First, when we add occlusion to the testing images, the performance drops dramatically. This is expected due to the information lost.
Second, when occlusion is applied, S-QAF outperforms the two baselines, \emph{i.e.}, feature concatenation and score-level averaging, indicating its ability to select and use the informative features. Third, when testing with original samples without occlusion, S-QAF is slightly inferior to the 
baselines. We examine the score curves and find that, in this case, 
all the component features tend to generate ``good curve'' (in which true matches are ranked at top). So the S-QAF module might generate less accurate estimation of feature importance. From this observation we speculate that, the proposed query adaptive fusion scheme 
would work at its prime if the to-be-fused features are not all ``good'' features. In such case the fusion scheme can assign 
reasonable weights to different features. 
This is also supported by the results in Section \ref{sec:exp:personrecognition}, where S-QAF 
is good at filtering out features of absence faces.

\textbf{Discussion}. Considering all the aforementioned experimental results, we find that  query adaptive fusion
works better if the following two conditions are satisfied. First, the features are complementary to each other. For a specific 
query, it is best if there exist both good features and bad features. If the features are all good or all bad, the importance 
estimation will be less accurate. Second, the feature has some decent discriminative ability, so it can generate a gap between the true matches and 
and false matches. The gap is utilized for feature importance estimation. 
\section{Conclusion}
\label{sec:conclusion}
In this article, we present query adaptive fusion for visual retrieval tasks. Our method leverages the 
ranking score curves to estimate the importance of heterogeneous features, and then fuses them. Specifically, for unsupervised tasks such as particular image retrieval, we propose an unsupervised fusion scheme (QAF); for  supervised tasks such as person recognition, we proposed a supervised method (S-QAF). Both methods are able to estimate the feature effectiveness for a given query. In this way, ``good'' features are up-weighted and ``bad'' features are down-weighted. In our experiment, we show the proposed method is robust to parameter changes, is superior to several other fusion methods, and is robust when many  ``bad'' features are present in the system. In the two tasks, \emph{i.e.,} particular object retrieval and person recognition, we demonstrate that QAF and S-QAF are able to produce very competitive accuracy compared with the state-of-the-art methods. The applicable scope and the limitation of the proposed method is also investigated in an occluded pedestrian retrieval experiment.
\ifCLASSOPTIONcaptionsoff
  \newpage
\fi



\bibliographystyle{IEEEtran}
\bibliography{reference.bib}
%



%

\begin{IEEEbiography}[{\includegraphics[width=1in,height=1.25in,clip,keepaspectratio]{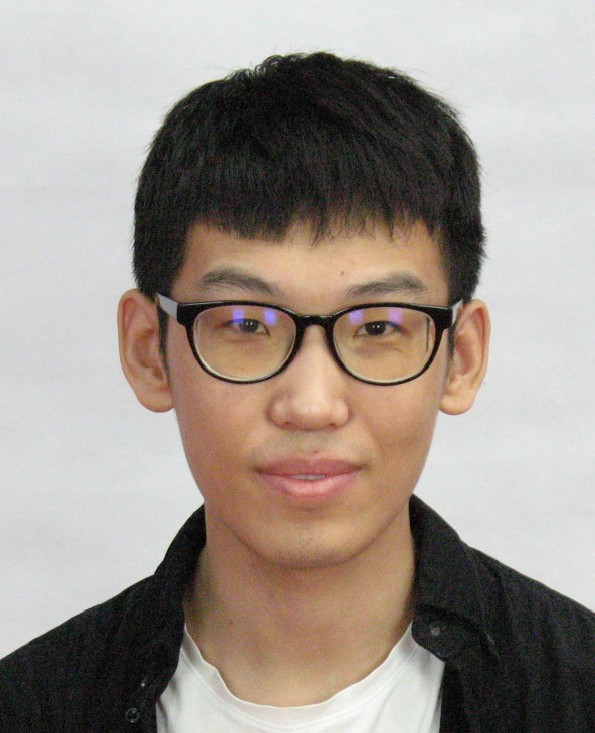}}]{Zhongdao Wang}
received his  B.S degree in the Department
of Physics at Tsinghua University in 2017. He is now
working towards the Ph.D. degree in the Department
of Electronic Engineering at Tsinghua University. His research 
interests include computer vision, pattern recognition and particularly 
 person/face recognition and retrieval.
\end{IEEEbiography}

\begin{IEEEbiography}[{\includegraphics[width=1in,height=1.25in,clip,keepaspectratio]{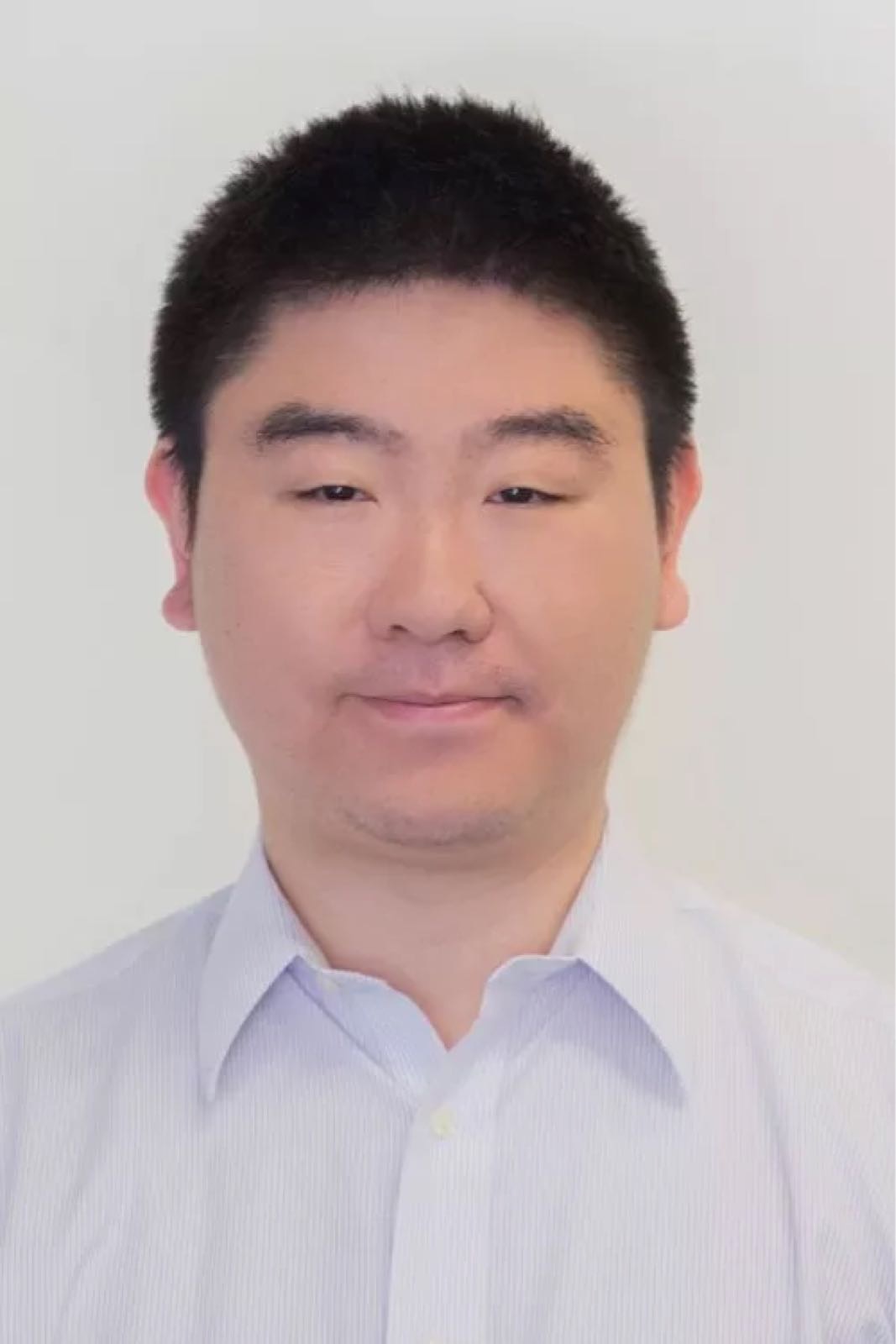}}]{Liang Zheng}
 is a Lecturer and a Computer Science Futures Fellow in the Research School of Computer Science, Australian National University. He received the Ph.D degree in Electronic Engineering from Tsinghua University, China, in 2015, and the  B.E. degree in Life Science from Tsinghua University, China, in 2010. He was a postdoc researcher in the Center for Artificial Intelligence, University of Technology Sydney, Australia. His research interests include image retrieval, classification, and person re-identification.
\end{IEEEbiography}


\begin{IEEEbiography}[{\includegraphics[width=1in,height=1.25in,clip,keepaspectratio]{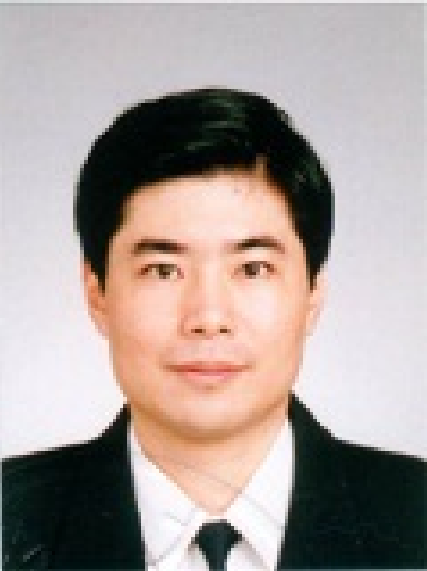}}]{Shengjin Wang}
received the B.E. degree
from Tsinghua University, China, in 1985 and
the Ph.D. degree from the Tokyo Institute
of Technology, Tokyo, Japan, in 1997. From
1997 to 2003, he was a member of Research
Staff in the Internet System Research Laboratories,
NEC Corporation, Japan. Since
2003, he has been a Professor with the
Department of Electronic Engineering, Tsinghua
University. He has published over
80 papers on image processing, computer
vision, and pattern recognition. 
His current research interests include image processing, computer
vision, video surveillance, and pattern recognition. He is a member
of the IEEE and the IEICE.

\end{IEEEbiography}




\end{document}